\documentclass[10pt,preprint]{aastex}
\singlespace
\shorttitle{850 Micron Survey of the HDF-N}
\shortauthors{Wang, Cowie, \& Barger}

\begin{document}

\title{An 850 Micron SCUBA Survey of the HDF-N GOODS Region}

\author{W.-H. Wang,\altaffilmark{1}
L. L. Cowie,\altaffilmark{1}
and A. J. Barger\altaffilmark{2,3,1}}
\altaffiltext{1}{Institute for Astronomy, University of Hawaii, 2680 Woodlawn Drive, Honolulu, HI 96822}
\altaffiltext{2}{Department of Astronomy, University of Wisconsin-Madison, 475 North Charter Street, Madison, WI 53706}
\altaffiltext{3}{Department of Physics and Astronomy, University of Hawaii, 2505 Correa Road, Honolulu, HI 96822}

\begin{abstract}
The Hubble Deep Field-North (HDF-N)  is one of the best studied extragalactic fields, 
and ultra-deep optical, radio, X-ray, and mid-infrared wide-field images are 
available for this area.  Here we present an 850 $\mu$m survey around the 
HDF-N, covering most of the area imaged by the Advanced Camera for Surveys 
as a part of the Great Observatories Origins Deep Survey.  Our map has 0.4-4 mJy
sensitivities (1 $\sigma$) over an area $\sim110$ arcmin$^2$ and there are 45 sources 
detected at $>3$ $\sigma$.  After correcting the effects of noise, confusion, 
incompleteness, and the Eddington bias using Monte Carlo simulations,
we find that the detected 850 $\mu$m sources with fluxes greater than 2 mJy have a
surface density of $3200^{+1900}_{-1000}$ deg$^{-2}$ and account for about 24\%
to 34\% of the far-infrared extragalactic background light.  Using the deep radio 
interferometric image and the deep X-ray image, we are able to accurately locate 
$\sim60\%$ of the bright submillimeter (submm) sources.  In addition, by 
assuming the Arp 220 spectral energy distribution in the submm and radio, we 
estimate millimetric redshifts for the radio detected submm sources, and redshift 
lower limits for the ones not detected in the radio.  Using the millimetric redshifts 
of the radio identified sources and spectroscopic and optical photometric redshifts
for galaxies around the submm positions, we find a median redshift of 2.0 for 11
possibly identified sources, or a lower limit of 2.4 for the median redshift of our $4$
$\sigma$ sample.  
\end{abstract}


\section{Introduction} 
Recent deep submillimeter (submm) surveys with the Submillimeter Common-User
Bolometer Array (SCUBA) on the James Clerk Maxwell Telescope (JCMT) and the 
Max-Plank Millimeter Bolometer array on the IRAM telescope have resolved the 
far-infrared (FIR) extragalactic background light (EBL) detected by COBE 
\citep[e.g.,][]{puget96,fixsen98,hauser98} at 850 $\mu$m and 1.2 mm into discrete
sources \citep{smail97,barger98,hughes98,barger99a,eales99,bertoldi00,chapman02,cowie02,scott02,eales03,webb03a}.  
In blank field surveys, the resolved point sources with 850
$\mu$m fluxes in the 2-10 mJy range account for 20\%-30\% of the FIR EBL.  
The 2-10 mJy flux range corresponds to $\sim10^{12-13}$ 
$L_{\sun}$ FIR luminosity at redshifts greater than 1, where the observed submm 
flux is not a function of redshift due to the steep dust spectrum in the submm 
\citep[see, e.g.,][]{blain93}.  Thus, the bright submm source population is 
inferred to be the distant analog of the local ultraluminous infrared galaxy 
population \citep{sanders96}.  In lensing cluster field surveys, a further 
45\% to 65\% of the FIR ELB is resolved into point sources with 0.3-2 mJy 
intrinsic 850 $\mu$m fluxes, corresponding to more ``normal'' $10^{11}$ 
$L_{\sun}$ galaxies (Cowie et al.\ 2002).

Observations suggest that these submm sources which dominate the FIR EBL are
mainly high redshift ($z>1$) starburst galaxies 
\citep{barger99b,barger00,fox02,chapman03b} whose submm emission mostly 
comes from dust heated by young massive stars.  Because the observed FIR EBL is
comparable with the total optical/UV EBL, these submm sources are crucial to
understand the integrated star formation history of the universe and may be the 
dominant component of star formation at $z>1$ 
(e.g., Barger et al.\ 2000; \citealp{gispert00,chapman03b}).

In order to better understand the evolution of the submm sources and the star
formation history, it is crucial to accurately determine the surface density of the
submm sources, to determine their properties at other wavelengths, and to measure their 
redshifts.  The surface density of submm sources brighter than 2 mJy was
previously determined to be only $\sim0.5$ arcmin$^{-2}$ 
(e.g., Barger et al.\ 1999a; \citealp{eales00}), implying that any statistical study of
this population requires a large survey area.  It is much more difficult, however, to
measure the redshift distribution of the submm sources.  The fundamental reason is 
that the current submm telescopes have very low resolution, as compared to optical 
telescopes.  In most cases, a few optical galaxies are found within a submm 
telescope beam, and it is difficult to unambiguously identify the real counterparts 
to the submm emission.  It is also time consuming to measure the redshifts for 
each of these counterpart candidates \citep{barger99b}.  Radio interferometric
imaging is the most commonly used method to solve this problem.  By assuming that
the correlation between the FIR and radio fluxes of normal galaxies (i.e., galaxies
without active galactic nuclei) in the local universe \citep[see, e.g.,][]{condon92}
also holds for the submm sources, radio sources near the submm positions could be
identified as counterparts (Barger et al.\ 2000), and the redshifts of the associated
optical galaxies may be measured \citep{chapman03b}.  In addition, redshifts of the
submm sources could also be independently estimated using the radio and submm
fluxes by assuming a plausible spectral energy distribution (SED, \citealp{carilli99};
Barger et al.\ 2000; \citealp{yun02}).
The limitation of the radio identifications is that only $\sim60\%$ of the 
\emph{bright} ($>6$ mJy) submm sources have radio counterparts (Barger et al.\
2000; \citealp{ivison02,chapman03a}).  Accurate positions and redshifts for most 
of the submm sources still cannot be measured and will have to await the advent
of high resolution submm observations which are now becoming possible with the
advent of the Submillimeter Array \citep{moran98}.

We have been carrying out an 850 $\mu$m SCUBA survey of intermediate depth 
(0.4 to 4 mJy 1 $\sigma$ sensitivity) over a large area ($\sim110$ arcmin$^2$) 
centered on the Hubble Deep Field-North (HDF-N).  Early results of this survey 
targeting optically faint radio sources were published in Barger et al.\ (2000).  
Our 850 $\mu$m survey covers most of the area imaged by GOODS (Great Observatories 
Origins Deep Survey) using the Advanced Camera for Surveys (ACS) on the 
\emph{Hubble Space Telescope} (\emph{HST}, \citealp{giavalisco04}) and a large part 
of the \emph{Chandra} Deep Field-North (CDF-N) observed by the 
\emph{Chandra X-Ray Observatory} (hereafter \emph{Chandra}; \citealp{alexander03}).  
This field also has deep radio and 
mid-infrared (MIR) imaging \citep{richards00,aussel99}, as well as ultra-deep
ground-based optical and near-infrared (NIR) imaging \citep{capak04a}.  The deep
radio image at 1.4 GHz provides accurate astrometry and millimetric redshift
estimates for the submm sources  that have radio counterparts.  Our goal is to better
constrain the number counts of bright submm sources, and to use
the deep multi wavelength data to understand the redshift distribution and the 
properties of the submm sources.  In this paper, we present the 850 $\mu$m source
catalog, number counts, and the optical, radio, X-ray, and MIR counterpart candidates
to the submm sources.

\section{Observations and Data Reduction}

SCUBA jiggle maps at 850 $\mu$m of the HDF-N flanking fields were obtained 
in multiple  runs between 1999 and 2003 under excellent submm weather 
conditions ($\tau_{225} < 0.08$).  The maps obtained in April and June 
1999 were published in Barger et al.\ (2000).  Each of the SCUBA maps covers 
an $\sim 2\farcm3$ field-of-view and has typical integration times between 
10 ks and 30 ks, depending on weather conditions and other observational 
constraints.  The maps were dithered with $10\arcsec$ to $20\arcsec$ 
offsets.  The dithering and the instrumental rotation on the sky prevent 
most regions of the sky from continuously falling on bad bolometers.  
Chopping of the secondary was fixed in the R.A. direction with a $45\arcsec$ 
chop throw.  Such a chop produces two $50\%$ negatives sidelobes to the east 
and west of the primary beam, which has a $14\farcs5$ 
full-width-half-maximum (FWHM).  The $45\arcsec$ chop throw was chosen so 
that each detected source would have at least one sidelobe inside the 
$2\farcm3$ field-of-view.  This effectively increases the on-source 
integration time.  

Pointing checks were performed before and after each 
$\sim 1$ hour of on-source observation and every time after transit.  
A nearby radio source 
($0954+685$, $1418+546$, $0923+392$, $1308+326$, or $1044+719$) 
was used for the pointing checks.  The typical pointing offset is 
$\lesssim 1\arcsec$, and no offset greater than $2\arcsec$ was observed.  
The JCMT has noted a newly discovered tracking 
error\footnote{http://www.jach.hawaii.edu/JACpublic/JCMT/Facility\_description/Pointing/tracking\_fault.html}.
We calculated the pointing error caused by this and found that most of our
target fields and pointing sources have combined tracking errors $\lesssim1\arcsec$, 
comparable to the normal pointing errors.  A few observations at coordinates 
$1237+6213$ have $1\farcs8$ errors, corresponding to 0.12 
beam FWHM.  These observations only contribute 25\% to the integration 
time in this region.  We thus conclude that our pointing and astrometry are
not seriously affected by this tracking error.

Flux calibration was done every night using the primary calibrator Mars or 
the secondary calibrators CRL 618, CRL 2688, IRC +10216, or OH 231.8.  If 
the variable calibrator IRC +10216 was used, a non-variable calibrator such 
as CRL 618 or a primary calibrator would be observed in the same run to 
confirm its light curve.  On each night, the flux was calibrated using a 
$30\arcsec$ aperture centered at the primary beam of the calibrator.  The 
size of the aperture is not critical in this research as long as the same 
aperture size is adopted each night.  This is because the filter function 
we used for flux measurements will be renormalized by a calibrator 
(see \S~\ref{id_cat}), and the flux measurement is independent of the 
aperture size.  The sky opacity was monitored during the observations in 
various ways.  Standard ``sky dips'' that give the most reliable 850 $\mu$m 
opacities were obtained every two to four hours, depending on the stability 
of the weather.  The JCMT water vapor monitor (WVM) was used when available 
to obtain the sky opacity every six seconds toward the same direction of the 
telescope beam.  When the WVM was not available, opacity values from the 
Caltech Submillimeter Observatory  Tau-Dipper ($\tau_{\rm CSO}$) were used 
to monitor the opacity change every ten minutes.  We found that for most of 
the time, the sky-dip values were consistent with the WVM values, while the 
$\tau_{\rm CSO}$ values have larger errors.

The jiggle maps described above contain 73.2 hours of integration in total.  
In addition to our maps, we also used that part of the ultra-deep jiggle map 
centered at the HDF-proper from the archive which had matched chopping.  
This ultra-deep map was first presented in \citet{hughes98} and was 
extensively analyzed by \citet{serjeant03}.  We did not attempt to fully 
reproduce their results, and we did not use all of the archival data.  We 
only included the data taken with a $45\arcsec$ east-west chop throw, 
identical to our standard one.  In other words, our final jiggle map 
contains all the SCUBA data taken with an identical strategy.   This gives 
a uniform point spread function (PSF) over the entire field and allows us 
to analyze the data in a consistent way (cf., the HDF super-map that 
contains a scan map and jiggle maps taken with various chops; 
\citealt{borys03}).  The data included here consist of 94.7 hours of 
integration and cover an area of $\sim110$ arcmin$^2$ with 0.4 to 4 mJy 
(1 $\sigma$) point-source sensitivity.

\begin{figure}[!h]
\epsscale{0.5}
\plotone{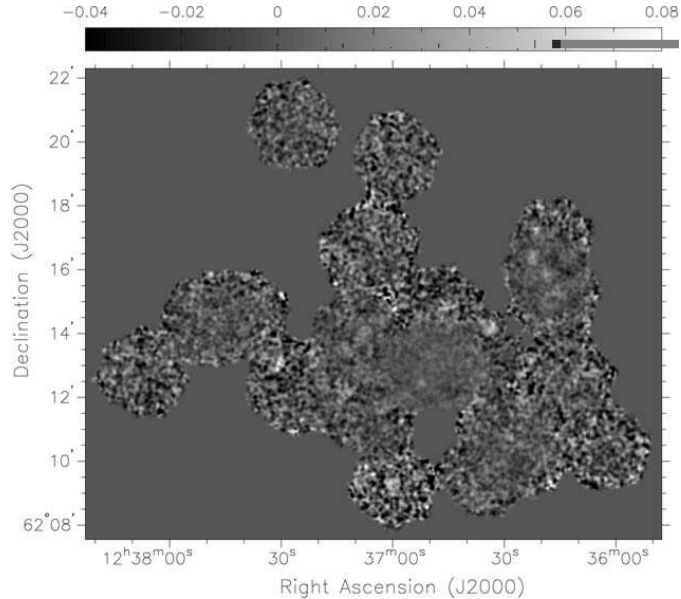}
\caption{850 $\mu$m SCUBA jiggle map of the HDF and flanking fields.   
The least noisy region close to the map center is the HDF-proper.  The 
grayscale of the map has units mJy arcsec$^{-2}$.  \label{hdf_sky}}
\end{figure}

The data were reduced using the package SCUBA User Reduction Facility 
(SURF; \citealp{jennes00}).  In SURF, the data were flat-fielded, 
atmospheric extinction corrected, and pointing corrected, and the sky 
noise was removed in standard ways\footnote{
\citet{borys03} mentions that in our spring 2003 data there is a periodic 
noise artifact with a timescale that is the same as the 16-point jiggle 
pattern.  We noticed this noise in our data taken in March 2003 during 
our data reduction.  About 2/3 of the bolometers suffered from this 
periodic noise (Borys, private communication) but only a few bolometers 
($<5$) showed strong noise signals.  We inspected all of the 
data and manually removed the bolometers showing such strong noise signals.  
Other bolometers potentially having this problem were excluded during the 
sky noise removal but were still included in our final maps.  Because we 
already took the bolometer variance (represented by the weight assigned 
to each bolometer) into account when making maps and extracting sources, 
these noisy bolometers should not affect our analyses.}.   
The extinction correction made use of the sky-dip results when the opacity 
was stable or the WVM results when the opacity varied rapidly.  Before maps 
were made, the data were weighted according to the bolometer variance 
relative to the central bolometer in the first observation.  Maps were made 
in SURF with the REBIN routine.  However, because REBIN cannot handle more 
than 200 observations (we had 244 observations), we divided the data into 
two roughly equal halves and REBINed them individually.  REBIN generated 
the sky maps, integration time maps, and weight maps from each half of the 
data.  We then combined the two sets of maps to form the final sky, 
integration time, and weight maps.  We present our final sky map in 
Figure~\ref{hdf_sky}. 
 
In addition to the sky map, we also constructed a ``true noise'' map in 
which all of the sources have been cleaned out, such that the map contains 
only the bolometer and sky noise, following the procedure introduced in 
Cowie et al.\ (2002).   After bright sources were detected and removed from 
each of the two half maps using the method described in \S~\ref{id_cat}, 
the true noise map was constructed by subtracting the two half maps from 
one another.  The map was then scaled by the factor 
$(t_1 t_2)^{1/2}/(t_1+t_2)$, where $t_1$ and $t_2$ are weighted integration 
times for each pixel in the two half maps.  This effectively removes all 
celestial objects.  Noise measured in this way is generally lower than that 
directly measured from the sky map, which contains confusion noise from 
undetected faint sources.  In particular, for the deepest region
at the HDF-proper where confusion becomes important, we found that 
the true noise is $\sim 10\%$ lower than the that measured from the cleaned 
sky map.  In this paper, all analyses are based on the true noise.

\begin{figure}[!h]
\plotone{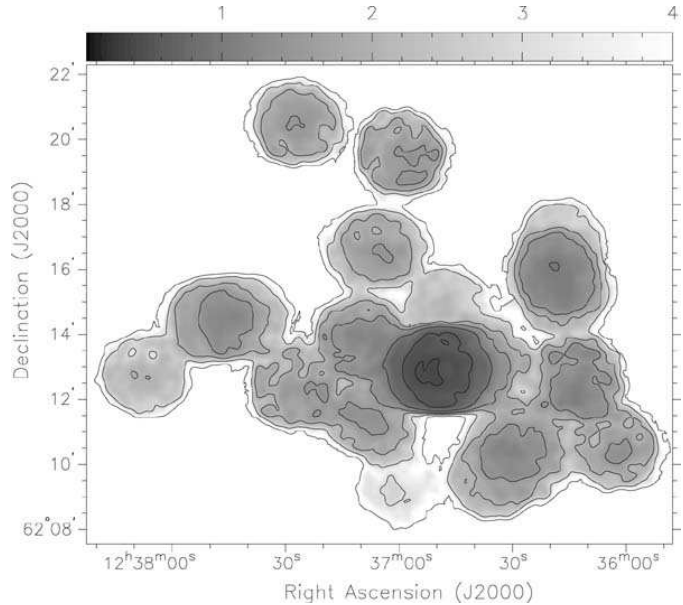}
\caption{Sensitivity (1 $\sigma$) distribution of the 850 $\mu$m SCUBA jiggle 
map based on the beam-optimized flux (see text).  From the innermost contour 
at the HDF-proper, the contour levels are 0.45, 0.6, 1.0, 1.5, 2.0, 3.0, and 
5.0 mJy.  The grayscale of the map has units mJy.  \label{hdf_rms}}
\end{figure}

\section{Source Identification and Catalogue}\label{id_cat}

Our source detection algorithm is optimized for single point-like sources.  
To detect such sources, the optimal filter function is the telescope beam 
inversely weighted by the noise in each pixel.   The telescope beam 
includes the primary beam and the two negative sidelobes.  This allows 
recovery of the integration times on the sidelobes and enhances the 
signal-to-noise ratio (S/N) of the detected sources.  Such S/N enhancement 
is up to $\sim 20\%$ if both sidelobes fall in the field-of-view.  The noise 
of each pixel is represented by the integration time and bolometer variance
that is contained in the weights assigned to the pixel.  We thus calculated 
the beam-optimized flux map ($F$, hereafter the flux map) using 
\begin{equation}\label{beamwflux}
F = \frac{(S \times T \times W)\otimes P}{(T \times W)\otimes P^2},
\end{equation}
where $S$ is the sky map, $T$ is the integration time map, $W$ is the weight 
map, $P$ is the normalized beam response, and $\otimes$ denotes convolution.  
The telescope beam response was obtained from observations of point-like 
calibrators.  It was normalized to yield correct point-source fluxes by 
using Eq.~\ref{beamwflux} on a calibrator.  The flux calculated with 
Eq.~\ref{beamwflux} also yields the minimum $\chi^2$ for point sources 
(see \citealt{serjeant03} for the derivation).  The error associated with 
the beam-optimized flux (i.e., 1 $\sigma$ sensitivity) was calculated by 
propagating the errors  in Eq.~\ref{beamwflux} using
\begin{equation}\label{beamwerror}
\sigma_F \propto \frac{1}{((T \times W)\otimes P^2)^{1/2}}.
\end{equation}
The proportionality is because SURF arbitrarily normalizes weight and 
integration time to the central pixel in the first observation.  The proper 
normalization of Eq.~\ref{beamwerror} utilizes the true noise map.  The 
error map from Eq.~\ref{beamwerror} was scaled by a constant so that the 
noise flux measured using Eq.~\ref{beamwflux} in the true noise map has a 
mean S/N of 1.  The normalized error map is presented in 
Figure~\ref{hdf_rms}.  A S/N map was made by dividing the flux map by the 
error map.  

\subsection{Direct Source Extraction}\label{direct_extract}

Source identification was performed in the S/N map.  S/N peaks calculated 
from Eq.~\ref{beamwflux} and Eq.~\ref{beamwerror} are peaks of minimum 
$\chi^2$.  In principal, these peaks could be considered as candidate 
detections and their fluxes and errors could be measured in the flux and 
error maps.  However, especially for faint sources, the fluxes measured 
in this way may be altered by the sidelobes of nearby sources.  There are 
several ways to solve this problem.  The simplest approach is to remove 
bright sources from the map before fluxes of fainter sources are measured.  
We note that the sources with high S/N are also the brighter ones locally.  
Therefore, we first measured the flux of the highest S/N peak and removed 
the corresponding PSF to 1.5 $\sigma$ from the sky map at the S/N peak 
location.  The amount of flux removed and the location of the S/N peak
were recorded.  This location of the S/N peak is used as the source 
position in the final catalog.  Then we recalculated the flux map 
and the S/N map from the 
residual sky map to identify and remove the next highest S/N peak.  We 
repeated this until the S/N was less than 3.0 everywhere.  The depth of 
source removal was 1.5 $\sigma$, i.e., there was 1.5 $\sigma$ residual 
flux on the sky map after each source was removed.  We observed that if 
values $\lesssim 1.0$ $\sigma$ (the flux uncertainty) were used, errors 
grew rapidly and some ``new sources'' would be created.  Because not all 
of the fluxes were removed, we measured the fluxes of all identified 
sources in the final residual flux map and added these fluxes back to the 
cataloged fluxes.  The procedure above reduces the interference of nearby 
sources to $<0.8$ $\sigma$ because the 1.5 $\sigma$ flux limit leaves a 0.8 
$\sigma$ residual flux in the sidelobes.  For sources that do not fall 
exactly on each other's sidelobes, the residual interference would be even 
smaller and thus negligible.  

\begin{figure}[!h]
\plotone{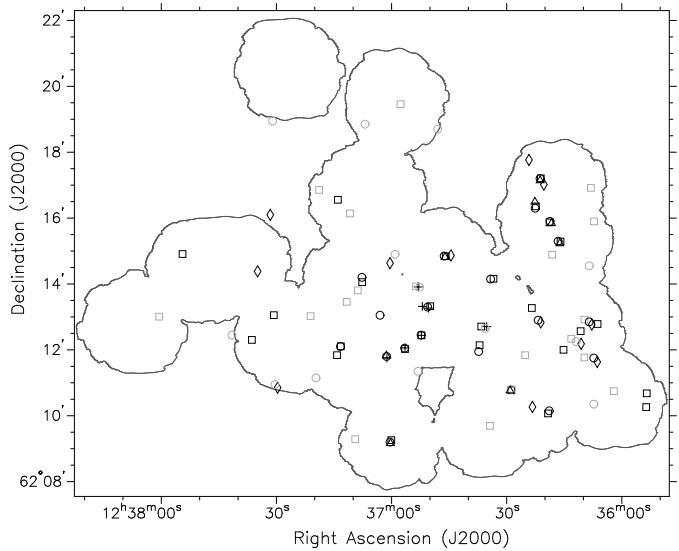}
\caption{850 $\mu$m sources around the HDF-N in various surveys.  Thick 
open squares are $>3.5$ $\sigma$ jiggle-map sources  in Table~\protect\ref{tab1} 
and thin open squares are 3.0-3.5 $\sigma$ ones.   Crosses are jiggle-map 
sources in \protect\citet{serjeant03}.  Triangles are jiggle-map sources 
in Barger et al.\ (2000) with $>2.7\sigma$ (all detected with $>4$ $\sigma$ 
here).   Diamonds are scan-map sources in \protect\citet{borys02}.  Thick 
open circles are $>4$ $\sigma$ super-map (jiggle map plus scan map) sources, 
and thin open circles are $>3.5$ $\sigma$ ones \protect\citep{borys03}.  
Note the major disagreements between the jiggle map sources and the scan 
map sources (see \S~\protect\ref{hdf_scan} and \protect\ref{hdf_super}).
\label{hdf_sources}}
\end{figure}

The source catalog constructed in this way is presented in Table~\ref{tab1} 
and Figure~\ref{hdf_sources}.  A total of 17 sources are detected at 
$>4 \sigma$.  Only sources detected above this significance level should 
be considered as secure detections.  At lower significance, many of the 
sources will be real, but some will correspond to noise peaks.  There are 
a further 28 sources between 3 $\sigma$ and 4 $\sigma$.  
In \S~\ref{n_counts} we estimated the number of spurious detections caused 
by noise to be $\sim7$ to 10 in the 3.0-4.0 $\sigma$ range, and most of 
these are in the 3.0-3.5 $\sigma$ range.  For the 10 sources in the 3.5-4.0 
$\sigma$ range we expect $\sim1$ is a spurious source.

The procedure described above assumes isolated point-like sources.  However, 
there might be extended sources or multiple blended sources which are 
marginally resolved.  It is also possible that the PSF of a detected source 
is damaged by a noisy bolometer.  For such sources, the PSFs are not 
point-like.  One example is GOODS 850-13.  After it was removed from 
the map, another significant peak (GOODS 850-23) still existed in its 
neighborhood.  Another example is source GOODS 850-6.  This source happened to 
fall close to a bad bolometer during the observations.  Its observed PSF is 
altered by that bolometer and shows two peaks (Figure~1), though still consistent with 
a single source.  Being aware of this, we also list fluxes measured within 
$30\arcsec$ diameter apertures in Table~\ref{tab1}.  Sources with aperture 
fluxes significantly greater than the beam-optimized fluxes 
(e.g., GOODS 850-5, 6, 13, and 16) may be marginally resolved by the telescope 
beam or affected by noisy bolometers.  We caution that for such sources, 
the cataloged multiples should not be considered as a unique configuration 
of the flux distribution.  The exact source shape and multiplicity can only 
be revealed by optical/near-IR or radio identifications, or by submillimeter 
interferometric imaging.

\subsection{CLEAN Deconvolution}

In addition to the direct source extraction described above, another method 
to solve the sidelobe interference is to deconvolve the map with the 
telescope beam, as most radio astronomers do.  We performed standard CLEAN 
deconvolution on our map, also based on beam-optimized fluxes.  In our CLEAN 
deconvolution, the highest S/N peak was identified and a small portion 
($10\%$ to $20\%$) of its flux was removed (CLEANed) from the sky map.  The 
next flux map was then constructed from the residual sky map and the next 
highest S/N peak was CLEANed, and so on.  Each time when some flux was 
CLEANed, the flux and the location were recorded in a catalog.  The CLEAN 
was stopped after the S/N on the map was $<1.5$ everywhere.  There is a 
difference between our CLEAN and the normal CLEAN used in most radio 
astronomy.  Our CLEAN removes sources at the beam-optimized S/N peaks while 
normal CLEAN removes sources at the peaks on the sky map.  Given the low S/N 
of the SCUBA map ($<5$ for most sources) and the very non-uniform
sensitivity, we believe that our method provides more reliable results.  

It is apparent that the direct extraction method in \S~\ref{direct_extract} 
is a simplified CLEAN---it removes the flux of a source at once, instead of 
just removing a small portion of the flux.  Indeed, the catalog generated by 
the CLEAN deconvolution is in excellent agreement with Table~\ref{tab1}, and 
there is no point in listing it separately.  The only differences between 
the CLEAN catalog and  Table~\ref{tab1} are in the marginally resolved 
sources.  For these sources, the direct extraction provided a configuration 
that has the fewest point-like components, while CLEAN generated many faint 
point sources within an area comparable to the beam size.  Again, we caution 
that for marginally resolved sources, both the direct extraction results and 
the CLEAN results are approximate models for the flux distribution.  Neither 
of them provides a unique deconvolution solution.

\section{Comparison with Previous 850 $\mu$m Catalogues}

The HDF and its flanking fields have been investigated at 850 $\mu$m with 
SCUBA by several other groups.   In this section, we compare our source 
catalog with the ones in these surveys.

\subsection{HDF-Proper Jiggle Map}\label{hdf_proper}

\citet{hughes98} and \citet{serjeant03} surveyed the HDF-proper area using 
the jiggle mode of SCUBA.  Their survey covered an area of 
$\sim10$ arcmin$^2$ with $\sim0.2$ to 1.5 mJy sensitivity. Our map of the 
HDF-proper region made use of a subset of the observations of \citet{hughes98} 
and \citet{serjeant03} but with independent data reduction.  We compared our 
catalog and the catalog in \citet{serjeant03} to look for systematic 
differences in flux calibration and astrometry.  Figure~\ref{hdf_sources} 
shows sources in the two catalogs.  Among the seven detected sources in 
Table~1 of \citet{serjeant03}, six were recovered by us, despite the fact 
that we only included $\sim1/3$ of the HDF-proper archival data.  We failed 
to detect their source HDF850.6 even though our map at its location is deeper 
than that of \citet{serjeant03}.  HDF850.7 was detected in our map, but with a 
$9\farcs9$ offset.  Because HDF850.6 and HDF850.7 are both at the map edge 
of \citet{serjeant03}, where the noise is significantly higher, the 
disagreement between the two catalogs on these sources is not unusual. 
Excluding HDF850.6 and HDF850.7, the mean positional offset between the two 
catalogs is $3\farcs8$ and the offsets appear to have random directions 
(see Figure~\ref{hdf_sources}).  The mean flux difference between the two 
catalogs is 4\%.  We conclude that our analysis is consistent with that of 
\citet{serjeant03} and that there is no systematic difference between our 
catalog and the \citet{serjeant03} catalog in terms of flux calibration and 
astrometry.  We note that HDF850.3 was detected by \citet{hughes98} but
not by \citet{serjeant03} and us.  HDF850.3 is located between the  
negative sidelobes of GOODS 850-10 and 19.  After GOODS 850-10 and 19 
were removed from the map, we found a very marginal signal ($1.49\pm0.50$ mJy) at the 
position of HDF850.3.  This flux is only 50\% of that in \citet{hughes98}.  
Although it is not qualified as a detection here, HDF850.3 may be a real source.

\subsection{Jiggle Maps in Barger et al.\ (2000)}\label{barger00_section}
Barger et al.\ (2000) published jiggle maps targeting optically faint 
microjansky radio sources in the flanking fields of the HDF 
\citep{richards00}.  Their reduced maps are included in this paper and 
re-analyzed.  There are a few differences between Barger et al.\ (2000) 
and this work.  Some regions in Barger et al.\ (2000) have been further deepened
after the publication.  The noise estimates of Barger et al.\ (2000) were based 
on CLEANed maps where confusion sources fainter than 3 $\sigma$ were not removed,
while in this work the noise estimates are free from faint sources.  Because
of these, sources detected here generally have smaller errors.  In Barger et al.\
(2000), submm fluxes were measured at radio positions, which are not necessarily
flux peaks in the submm map.
There are eight sources with 850 $\mu$m S/N greater than 2.7 cataloged in 
Barger et al.\ (2000, see Figure~\ref{hdf_sources}).  All of them are 
detected in this work with S/N greater than 4.0, mostly because of the 
increase in depth.  For these eight sources, the fluxes are in good agreement and the 
mean flux ratio (Barger et al.\ 2000 to this work) is $0.90\pm0.13$.  The slight
flux boost in this work can be explained by the fact that the submm fluxes
were measured at flux peaks instead of the radio positions.  

\subsection{HDF Flanking Field Scan Map}\label{hdf_scan}
 \citet{borys02} used the scan mode of SCUBA to survey a 125 arcmin$^2$ area 
with $\sim3$ mJy 1 $\sigma$ sensitivity in the flanking fields of the HDF.  
Our jiggle map has a survey area comparable to that of the scan map but with 
sensitivity up to 3 times higher.  The difference between our jiggle map 
and the scan map of \citet{borys02} is quite large, even at the 4 $\sigma$ 
level.   All 12 scan-map sources have fluxes $>10$ mJy, and ten are well 
inside our survey area (see Figure~\ref{hdf_sources}).  However, only three 
scan-map sources in Table~1 of \citet{borys02} were detected by our jiggle 
map and these had fairly large positional offsets.  They are 
HDFSMM--3608$+$1246 (associated with our source GOODS 850-8), 3620$+$1701 
(GOODS 850-15), and 3644$+$1452 (GOODS 850-11).  The offsets between the 
scan-map and the jiggle-map source positions are $10\farcs5$, $12\farcs6$, and 
$9\farcs9$, respectively.  The offsets seem to have random directions (see 
Figure~\ref{hdf_sources}).  On average, the scan-map fluxes of the three 
sources are $42\%$ greater than the jiggle-map fluxes.

We measured the jiggle-map fluxes and the associated errors at the scan-map 
source positions.  The results are summarized in 
Table~\ref{tab2}.  Only the three sources mentioned above were detected at 
$>3$ $\sigma$.  We also looked for jiggle-map sources within $30\arcsec$ 
($\sim2\times$ the beam FWHM) centered at the scan-map source positions.  
As shown in Table~\ref{tab2}, except for the three detections, all 
jiggle-map sources found have either fluxes that are too low 
($\lesssim 1/3$ of the scan-map fluxes) or offsets that are too large to 
be associated with the scan-map sources.  We conclude that only three of 
the ten scan-map sources within our field-of-view were detected in the 
jiggle map.  This is a surprising result because our jiggle map is 
considerably deeper than the scan map.  If the scan-map sources were all 
real, then our jiggle map should have detected most of them.

Furthermore, we note that two very bright jiggle-map sources were not 
detected by the scan-map.  They are GOODS 850-6 and 16.  GOODS 850-6 has 
a jiggle-map flux of 13.6 mJy at $6.01$ $\sigma$.  There is a $107.0 \pm9.6$ 
$\mu$Jy 1.4 GHz source \citep{richards00} associated with this source.   Source
GOODS 850-16 has a jiggle-map flux of 12.45 mJy at $4.32$ $\sigma$.  There 
is a $324.0\pm18.0$ $\mu$Jy 1.4 GHz source \citep{richards00} associated 
with this source.  Therefore, in terms of the 850 $\mu$m S/N ratios and the 
FIR--radio correlation, these two jiggle-map sources are highly significant.  Both
sources are within the scan-map field-of-view, but neither are detected by the 
scan-map.  Thus, we conclude  that there is a major discrepancy between our jiggle
map and the scan map of \citet{borys02}.

\subsection{HDF Super-Map}\label{hdf_super}
\citet{borys03} constructed an ``HDF super-map'' by combining the scan map 
of \citet{borys02}, all the jiggle maps from \citet{serjeant03} , and a 
subset of our jiggle maps described here, which were already in the JCMT 
archive.  Given the very serious disagreement between the scan map and the 
jiggle map, it is not obvious to us what the combination of these two maps 
actually means.  Nevertheless, we briefly summarize the similarity and 
difference between the catalogs. 

Among the first 19 4 $\sigma$ sources in \citet{borys03}, 14 are detected 
($>3\sigma$) in our jiggle map with  positional offsets less than one beam 
FWHM.  The mean super-map-to-jiggle-map flux ratio of these 14 sources is 
$0.96\pm0.19$.  Of the remaining undetected sources, one is HDF850.6, 
mentioned in \S~\ref{hdf_proper}, and two are associated with the undetected 
bright scan-map sources HDFSMM--3606+1138 and 3621+1250 
(see Table~\ref{tab2}).  We extended this comparison to the full list of 3.5 
$\sigma$ sources in \citet{borys03}.  Among the 34  sources, only 17 are 
detected in our jiggle map.  We noticed that  bright sources below 4 
$\sigma$ are over-populated in their catalog.  Five sources have fluxes 
$\gtrsim20$, mJy while there are no $>4$ $\sigma$ sources with comparable 
fluxes.  All of these five bright sources are either outside or at the edge 
of our jiggle map, i.e., they come from the scan map.  This further enhances 
our conclusion in \S~\ref{hdf_scan} --- the scan map is inconsistent with our 
analysis.

\subsection{Photometry Observations}
\citet{chapman01} measured 850 $\mu$m fluxes of optically faint radio sources
in the HDF with the photometry mode of SCUBA.  Two of their targets fall in
our jiggle-map field-of-view.   The source VLA J123606+621021 has a photometry
flux of $11.6\pm3.5$ mJy and a jiggle-map flux of $2.21\pm1.68$ mJy.  This source
is detected at 3.7 $\sigma$ in the super-map but not detected in the scan map.  
The source VLA J123711+621331 is close to our GOODS 850-36 with a $4\farcs2$ 
offset.  It has a photometry flux of $7.7\pm2.4$ mJy, and a jiggle-map flux of 
$3.98\pm1.33$ mJy at the radio position or $4.4\pm1.36$ mJy if the flux of HDF 
850-36 is adopted.  For this source, although the detections agree with each other,
there is an up to 90\% difference between the photometry and jiggle-map fluxes.  
This source is not detected in either the scan map or the super-map.  Because the two
photometry detections and the jiggle-map source GOODS 850-36 are all marginal
(S/N $\sim3.0$), the photometry results and our jiggle map results are only
marginally inconsistent.    However, the large overestimates of the 
photometry fluxes in these cases do illustrate the danger of using 3 $\sigma$
detections in targeted photometry measurements, where one only integrates to a fixed
S/N.  Such a procedure will always result in overestimating the fluxes and should
be avoided.  

Radio source VLA J123600.2+621047 was listed in \citet{chapman03b} but
no details about its observation were explained in \citet{chapman03b} and 
previous papers.  We found that this source was observed in photometry mode, 
according to the SCUBA archive.  This source 
has a $7.9\pm2.4$ mJy photometry flux \citep{chapman03b}.  Our jiggle-map flux
at the radio position is $0.35\pm1.5$ mJy.  We notice that it is $6\farcs4$ 
away from the negative sidelobe of GOODS 850-25.  Our non-detection 
of VLA J123600.2+621047 might be due to the sidelobe of GOODS 850-25.  
However, to fully cancel a 7.9 mJy flux, GOODS 850-25 has to
be at least 15.8 mJy, which is unlikely.  This suggests that the 7.9 mJy 
photometry flux may be also an overestimate.   Unfortunately, the western sidelobe
of GOODS 850-25 is outside out map and the eastern sidelobe of 
VLA J123600.2+621047
(-3.9 mJy, assuming the photometry flux) is below our sensitivity limit.  We cannot 
unambiguously determine their fluxes using our jiggle map alone.

\section{Monte Carlo Simulations}\label{simulation}
\subsection{Number Counts}\label{n_counts}

\begin{figure}[!h]
\epsscale{0.4}
\plotone{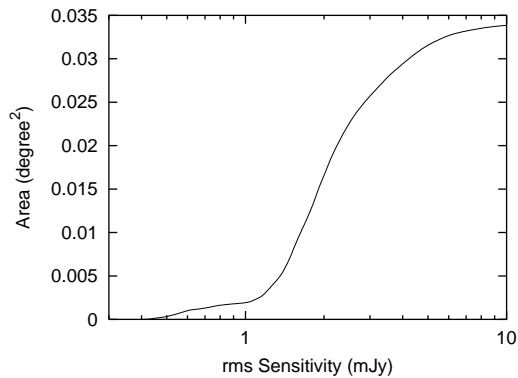}
\caption{Cumulative survey survey with 1 $\sigma$ sensitivity.  
The tail at $<1$ mJy is from the HDF-proper area.  It allows us 
to determine the number counts to the 2  mJy level.
\label{area_curve}}
\end{figure}

The differential number counts ($N(S)$) are determined by dividing the 
number of detected sources in some flux ($S$) interval by the area over 
which these sources could be detected.  The differential counts of our 
sample can be constructed using the source catalog and the survey area 
summarized in Figure~\ref{area_curve}.  For relatively small samples, 
cumulative number counts ($N(>S)$), which are integrals of the differential 
counts, are more commonly adopted.  The caution for cumulative counts is 
that the points are not statistically independent.  Errors in the bright 
end of the cumulative counts propagate to the faint end and changes in 
the shape of the counts may be hard to see.

\begin{figure}[!h]
\epsscale{0.45}
\plotone{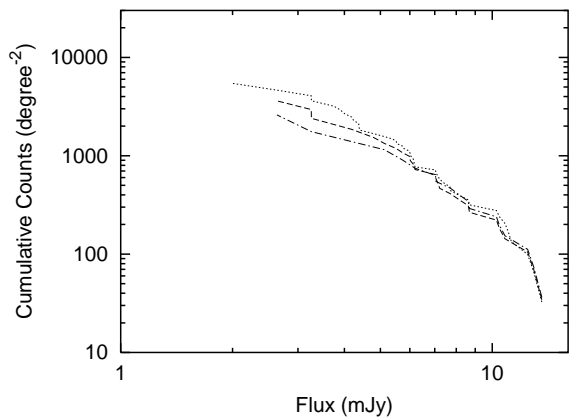}
\caption{Cumulative 850 $\mu$m number counts derived from our 3.0 $\sigma$ 
(dotted line), 3.5 $\sigma$ (dashed line), and 4.0 $\sigma$ sources 
(dash-dotted line).    The raw counts contain complex effects of noise, 
confusion, and systematic biases.  Although the 3.0 $\sigma$ counts are 
affected by noise more seriously, all the 3.0, 3.5, and 4.0 $\sigma$ counts 
are consistent with a single power law (\S~\protect\ref{simulation}).
\label{raw_counts}}
\end{figure}

\begin{figure}[!h]
\plotone{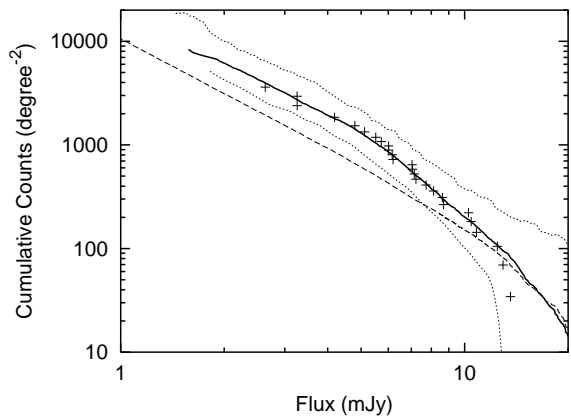}
\caption{Results of Monte Carlo simulations.  Observed counts (crosses) 
are from the 3.5 $\sigma$ sources.  Each cross in the plot represents a 
detected source in Table~\protect\ref{tab1}.  Solid line is the averaged 
number counts derived from the simulations that contain 100 realizations 
and $\sim3100$ 3.5 $\sigma$ sources.  Dotted lines are the 90\% confidence 
range for the observed counts derived from each realization.  Dashed line 
is the input power-law counts described in  Eq.~\protect\ref{fit_power_law}.  
The input counts start to turn over at $>10$ mJy because of the 25 mJy 
upper cutoff used in the simulations.  Because of the large uncertainties 
at $>10$ mJy, the upper cutoff is not constrained by the observations.
\label{sim_counts}}
\end{figure}

In Figure~\ref{raw_counts} we present raw cumulative 850 $\mu$m counts, 
derived from our 3.0 $\sigma$, 3.5 $\sigma$, and 4.0 $\sigma$ sources.  We 
note that sources detected at lower significance levels are not necessarily 
fainter in flux because our map sensitivities are highly nonuniform.  Raw counts 
constructed from lower significance samples are systematically higher in the plot
because of spurious sources caused by noise and because of Eddington bias.  
At the 3.0 and 3.5 $\sigma$ levels, the Gaussian probabilities of spurious 
detections are $4.4\times10^{-3}$ and $8.7\times10^{-4}$, respectively.  
Given the $\sim110$ arcmin$^2$ survey area, which corresponds to $\sim1600$ 
primary beams (i.e., $\sim1600$ independent points), we estimated the 
numbers of spurious sources in our 3.0 and 3.5 $\sigma$ samples to be 
$\sim7$ and $\sim1$, respectively.  We note that the Gaussian assumption 
is supported by the noise analysis in Cowie et al.\ (2002), and the above result 
is also consistent with our true noise map.  We used our source extraction 
algorithm to find ``sources'' in the true noise map.  The number of spurious 
sources found from the true noise map above 3.0 $\sigma$ is 10, consistent 
with the estimate based on Gaussian noise.  While the spurious sources 
mainly affect  the 3.0-3.5 $\sigma$ level, the Eddington bias affects the 
number counts at all S/N levels.  We will show that the differences between 
the 3.0 $\sigma$, 3.5 $\sigma$, and 4 $\sigma$ counts are indeed consistent 
with each other if the effects of noise, confusion, and systematic biases, 
such as incompleteness and the Eddington bias, are all taken into account.  
To do this, we performed iterative Monte Carlo simulations to estimate these 
effects and to derive the number counts.  

We created simulated images by randomly drawing sources from a plausible 
power-law number count relation onto the true noise map.   We limited the fluxes 
of the input sources to be between 0.2 and 25 mJy.  The deepest region in 
our map has a $0.4$ mJy rms sensitivity, so only sources brighter than 1.2 
mJy can be detected in this region.  Thus the 0.2 mJy lower cutoff is 
sufficient for our purpose and is also consistent with the fact that most 
of the 850 $\mu$m EBL arises in sources brighter than 0.3 mJy 
(Cowie et al.\ 2002).  The upper cutoff produces a turn-over on the bright-end 
counts (see Figure~\ref{sim_counts}).  However, the 25 mJy value used here 
is not important because there are not enough sources at this flux level 
to tightly constrain the upper cutoff.

We used the procedures described in \S~\ref{direct_extract} to detect the 
simulated sources.  To derive the averaged output counts, we ran 100 
realizations over the whole field (corresponding to $\sim3.1$ deg$^2$) and 
detected $\sim3100$ simulated sources at 3.5 $\sigma$.  We compared the 
recovered 3.5 $\sigma$ counts and the input power law to derive the bias 
with flux and used this to correct the observed 3.5 $\sigma$ counts.  We 
fitted the corrected observed counts with a power law using the area-weighted 
maximum likelihood method \citep{crawford70} to account for the statistical 
interdependence of the points in the cumulative counts.  We used this 
fitted power law as the input for the next simulations.  We repeated this 
process until the fitted power law from the observations matched the input 
power law within the fitting errors.  The final number counts in the 2--10 
mJy range determined by the power law fit are
\begin{equation}\label{fit_power_law}
N(>S) = 1.09  \times 10^4 (S/ \rm mJy)^{-1.76} ~~deg^{-2}.
\end{equation}
We present in Figure~\ref{sim_counts} this power-law input counts (with 25 
mJy upper cutoff), the averaged output counts from the simulations, and the 
3.5 $\sigma$ raw counts from the observations.  The 90\% confidence range of the 
observed counts is obtained by measuring the spread of the output counts 
over the various realizations and is presented in Figure~\ref{sim_counts}.  
For a convenient reference, we approximate the bias-corrected upper and lower
90\% confidence ranges between 2 and 10 mJy with the power laws
$N(>S) = 1.85 \times 10^4 (S/ \mathrm{mJy)^{-1.85} ~~deg^{-2}}$ and 
$N(>S) = 7.3 \times 10^3 (S/ \rm mJy)^{-1.75} ~~deg^{-2}$, respectively.

As mentioned previously, the 3.0 $\sigma$ raw counts are affected by 
spurious sources.  However, this effect is taken into account by the use 
of the true noise map.  Running similar simulations on the 3.0 $\sigma$ 
and 4.0 $\sigma$ sources provided fitted power laws similar to that of 
Eq.~\ref{fit_power_law}, and the differences between these counts are well 
within the uncertainties.   Although we used the 3.5 $\sigma$ results in 
this paper, we note that the counts derived from the 3.0 $\sigma$, 3.5 $\sigma$, 
and 4.0 $\sigma$ samples are all consistent with each other.

As shown in Figure~\ref{sim_counts}, the counts are best determined in 
the 2--10 mJy range.  The larger uncertainties at the bright and the faint 
ends are caused by the small numbers of sources detected at these flux 
ranges.  Between 2.0 and 10 mJy, the recovered counts exceed the input 
counts due to the Eddington bias.  Detections at a given flux range include 
fainter sources with fluxes boosted by positive noise, as well as brighter 
sources dimmed by negative noise.  For power-law distributions with negative 
indices, there are more flux boosted faint sources compared to dimmed bright 
sources.  This will cause a positive systematic flux boost and an upward 
shift in the number counts.  Over the 2.0-10 mJy range, we measured the 
median flux boost from the recovered counts to be  44\% for the 3.5 $\sigma$ 
counts.  \citet{eales00} found a median flux boost factor of 44\% for their 
sources brighter than 3 mJy, and \citet{scott02} found boost factors of 28\% 
and 35\% at $>5$ mJy for their two areas, both of which are consistent with 
the present analysis.

At 5 and 10 mJy, the bias-corrected cumulative counts 
(Eq.~\ref{fit_power_law} and the dashed line in Figure~\ref{sim_counts}) are 
respectively $640^{+300}_{-200}$ and $190^{+70}_{-60}$ deg$^{-2}$, where the 
uncertainties are derived from the simulated 90\% confidence ranges.  This 
is consistent with the \citet{scott02} measurements of $620^{+110}_{-190}$ 
and $180\pm60$ deg$^{-2}$ at 5 and 10 mJy, respectively, the Eales et al.\ (2000) 
measurement of $500\pm200$ deg$^{-2}$ at 5 mJy, and the Barger et al.\ (1999a)
measurement of $610^{+240}_{-190}$ deg$^{-2}$ at 5 mJy.  At the fainter 2 
mJy end, our cumulative counts are $3220^{+1910}_{-1050}$ deg$^{-2}$.  The 
other determinations at 2 mJy are $3500^{+1500}_{-1000}$ (Cowie et al.\ 2002), 
$2900\pm1000$ \citep{smail02}, and $6800^{+2600}_{-1900}$ deg$^{-2}$ 
\citep{chapman02}.  Our counts are consistent with these values within the 
errors.

The total surface brightness of submm sources can be derived from the number 
counts.  Using the corrected counts in Eq.~\ref{fit_power_law} and the 90\%
uncertainties, we found that the contribution to the 850 $\mu$m EBL in the 2--10 mJy
flux range is $1.05^{+0.62}_{-0.34}\times10^4$ mJy deg$^{-2}$.  The percentage of
the 850 $\mu$m EBL residing in this range is thus $34^{+20}_{-11}\%$, if we adopt
the 850 $\mu$m EBL measurement of $3.1\times10^4$ mJy deg$^{-2}$ from 
\citet{puget96}, or $24^{+14}_{-7}\%$, if we adopt the measurement of 
$4.4\times10^4$ mJy deg$^{-2}$ from \citet{fixsen98}.  Eales et al.\ (2000) and
Webb et al.\ (2003a) adopted the Fixen et al.\ (1998) value and found the fraction
of resolved  850 $\mu$m EBL to be 19\% and 13\%, respectively, for bias-corrected 
source fluxes greater than 2 mJy.  Our value of 24\% is consistent with the values
of the above two groups within the error.

\subsection{Reliability of Source Extraction}\label{reliability}

\begin{figure}[!hb]
\epsscale{0.7}
\plottwo{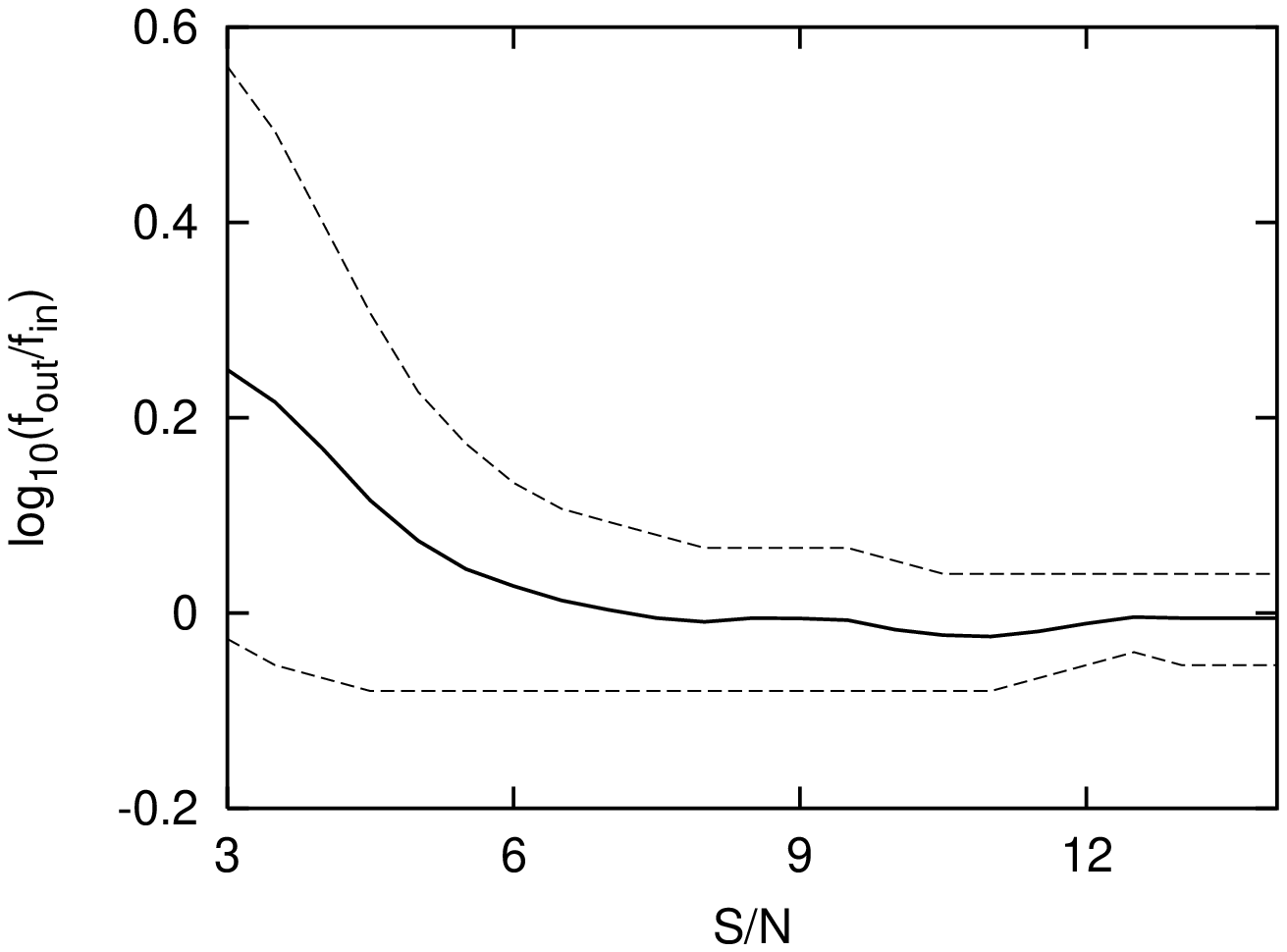}{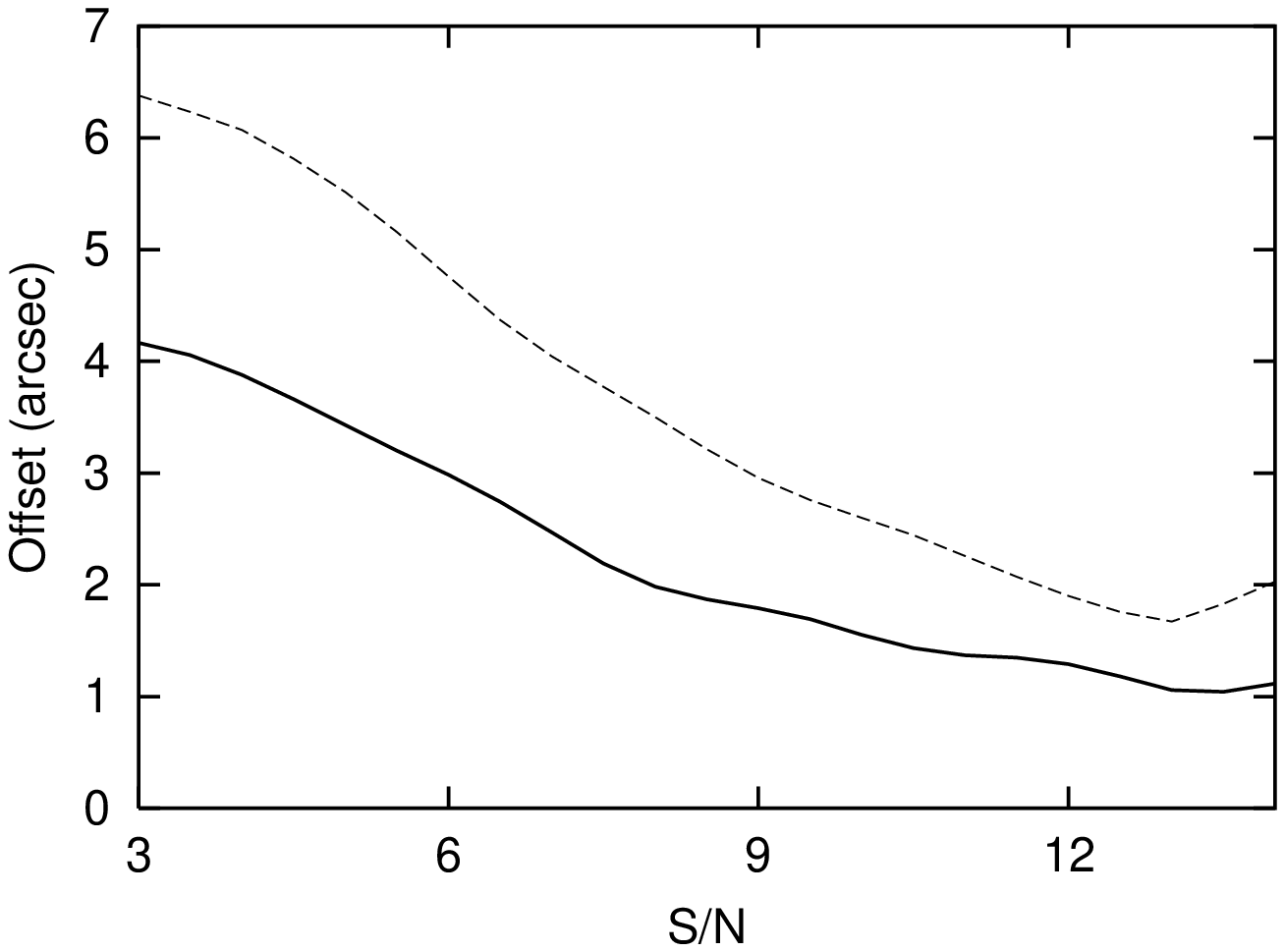}
\caption{Output to input flux ratio and positional error of our extraction 
versus S/N, as derived from our Monte Carlo simulations.  Solid lines show 
the mean flux ratios and the mean positional offsets.  Dashed lines are 
$\pm1\sigma$ flux ratios and offsets. In the offset plot, only the $+1\sigma$ 
line is shown and 84\% of the detections have positional errors below the 
$+1\sigma$ line.  The offsets presented here only account for the effects 
of noise and confusion.  Telescope pointing error and astrometry error are 
considered in \S~\protect\ref{counterparts}. \label{fratio_offsets}}
\end{figure}

Using the Monte Carlo simulations above, we studied the reliability of our 
source extraction.  Here we focus on source flux, spurious detections, and 
positional error.  Within one beam around each detected source in a  
simulation, we search for input sources and calculate the total flux 
contributed by these input sources. We plot the mean output (detected) to 
input flux ratios in Figure~\ref{fratio_offsets} versus S/N (solid line).  
To show the spread of the flux ratios, in addition to the mean flux ratios, 
we also plot the $\pm1\sigma$ flux ratios (dashed lines).  As discussed 
previously (the Eddington bias), because there are more fainter sources, 
the mean flux ratios are always greater than 1.  For the power-law number 
counts in Eq.~\ref{fit_power_law}, the mean flux ratio at 4 $\sigma$ is 
1.47, and the median for $>3.5\sigma$ in the 2-10 mJy range is 1.44.  We 
note that we did not find the flux ratios to be a strong function of flux 
(as opposed to S/N) because our sensitivities are highly nonuniform.  For 
the same reason, we did not attempt to quantify the completeness with flux.  
The effect of completeness is already included in Figure~\ref{sim_counts} 
and in the derivation of Eq.~\ref{fit_power_law} and the EBL contribution
of the sources.

To derive positional errors, we first excluded detections with large 
output-to-input flux ratios.  If the total input flux contributed less 
than 40\% of the detected flux, we considered the detection to be spurious.  
There are on average 10.6 spurious detections in each realization, and 9.0 
of them are $<4$ $\sigma$ detections.  We note that the choice of 40\% is 
arbitrary because the distribution of source flux and noise flux are both 
continuous.  The number of spurious detections defined in this way is very 
slightly larger than the Gaussian analysis and consistent with the true 
noise map in the previous section.  Among the real detections where the 
input fluxes contributed more than 40\% of the detected fluxes, there was 
usually more than one input source found within the beam.  Most of the input 
sources are faint and have fluxes less than 1 mJy.  In 83\% of the real 
detections, one input source dominates the total input flux at the $>50\%$ 
level.  We used the brightest input source within the beam to calculate  
positional error, whether this brightest source dominated the total input 
flux or not.  We plot the mean positional offsets versus S/N in 
Figure~\ref{fratio_offsets} (solid line).  To show the distribution of the 
offsets for a given S/N, in addition to the mean offset, we also plot the  
+1 $\sigma$ of the offset distribution (dashed line), i.e., 84\% of the 
detections have offsets below the +1 $\sigma$ line.  For S/N less than 6, 
while the mean offset is fairly small (3-4$\arcsec$, corresponding to 
$\sim1/4$ of the beam FWHM), $\gtrsim15\%$ of detections have offsets 
greater than $5\arcsec$.   

We note that the positional error is a consequence of noise and confusion.  
Therefore, the offsets are expected to be a complex function of flux and 
S/N.  The use of the true noise map in our simulations should provide a 
reasonable estimate of the effect of the noise.  The effect of confusion 
is simulated by our input power law in Eq.~\ref{fit_power_law}, which is 
consistent with the observations.  However, because the sensitivity is 
highly nonuniform across our map, we only saw the mean offsets to be a 
function of S/N, not as a function of flux.  We also point out that the 
positional error presented here only accounts for the effect of noise and 
confusion.  We can use the radio positions to test for other effects, such as the 
pointing error of the telescope and astrometry error (see the next section).

\section{1.4 GHz, X-ray, and Mid-Infrared Counterpart Candidates}\label{counterparts}
  
In this section we describe the 1.4 GHz and X-ray sources that are identified as
\emph{candidate} counterparts to the submm sources.  The fundamental difficulty of
identifying counterparts to the submm sources is the low resolution and the low S/N
in the submm observations.  In \S~\ref{reliability} we showed that our source
extraction could have positional errors ($+1\sigma$) between $2\arcsec$ and
$6\arcsec$, depending on the S/N.  To estimate telescope pointing and astrometry
errors, we offset the submm map by $0\arcsec$ to $10\arcsec$ to maximize the mean
submm fluxes of 1.4 GHz sources.  We found that the maximum mean submm fluxes
were measured with $1\arcsec$ offsets in both the right ascension and declination
directions.  While these offsets indicate a small absolute value for the general pointing
and astrometry errors for the whole map, the errors for individual jiggle maps could
be larger.  To account for the pointing and astrometry errors, we added a $2\arcsec$
rms error in quadrature onto the $+1\sigma$ offsets shown in Figure~\ref{fratio_offsets} 
and calculated the positional error for each 
detected submm source (Table~\ref{tab1}).  The error circles are shown in
Figure~\ref{thumbnails} and can be used to judge how likely a
radio or X-ray source is the counterpart to the submm source.  Approximately 84\% 
of the real counterparts should fall in the error circles.  Hereafter in this paper, we
shall restrict most of our discussion to $>4\sigma$ sources because of their relatively
small positional errors and more secure detections.

\subsection{1.4 GHz Counterpart Candidates and Millimetric Redshifts}\label{radio_id}

There is a tight correlation between the 1.4 GHz fluxes and FIR fluxes of 
normal galaxies in the local universe \citep[see, e.g., ][]{condon92}.  For 
high-redshift submm sources, the ratio of the submm to 1.4 GHz flux is an increasing
function of redshift (\citealp{carilli99}; Barger et al.\ 2000; \citealp{yun02}) because
the thermal dust emission in the submm and the nonthermal
emission in the radio have different spectral slopes.  However, we still anticipate 
that a fraction of the brighter submm sources will have radio counterparts according
to the FIR--radio correlation.  For sources detected at both submm and radio 
wavelengths, their redshifts can be estimated.  In addition, the radio 
detection provides subarcsec astrometry for the submm sources because of 
the high resolution of radio interferometers, allowing us to identify 
the optical counterparts and make spectroscopic observations.

\citet{richards00} presented a catalog of 372 1.4 GHz sources detected in a
Very Large Array map centered on the HDF-N that covers a $40\arcmin$ diameter
region with an effective resolution of $1\farcs8$.  The 5 $\sigma$ completeness limit
for compact sources in the central region is 40 $\mu$Jy.  Eighty seven of the sources
are within our submm field-of-view (850 $\mu$m rms sensitivity $<4.0$ mJy).  We
measured the submm fluxes at the positions of these radio sources and looked for 3
$\sigma$ detections.  Before we measured the submm flux at a radio position, we
removed bright submm sources in Table~\ref{tab1} that are at least one beam away
from the radio position.  This minimizes the sidelobe interference of nearby bright
submm sources.  In this way, 14 radio sources were found to have $>3\sigma$ submm
detections.  The eight submm detected radio sources in Barger et al.\ (2000) are all
recovered here.  Thirteen of the 14 submm detected radio sources are associated with
12 (out of the 17)
$>4\sigma$ submm sources; two of the radio sources are associated with one submm
source (GOODS 850-11).  There are 13 submm sources in Table~\ref{tab1} with
$S_{850}$ $>6$ mJy and S/N $>4$.  Eight, or $\sim60\%$, have radio sources
associated with them.  This fraction of radio detections agrees with other analyses of 
bright submm sources \citep[Barger et al.\ 2000;][]{ivison02,chapman03a}.  The
offsets between the radio positions and the beam-optimized submm positions are all 
within $9\arcsec$, with a vector mean offset of $1\farcs38$.  Nine of the 13 radio 
sources have offsets less than or comparable to the positional uncertainties of the 
submm sources.  We summarize these radio sources in Table~\ref{tab3} and 
Figure~\ref{thumbnails}.

In order to understand the fraction of random 1.4 GHz sources in these nine 
counterpart candidates, we made estimates based on the surface density of 1.4 GHz
sources and Monte Carlo simulations.  Using the 40--1000 $\mu$Jy number
counts in \citet{richards00}, we calculated that 0.019 and 0.043 1.4 GHz sources 
will be found within a $6\arcsec$  (roughly corresponding to our largest 
error circle) and a $9\arcsec$ circle,  respectively, if the distribution 
of 1.4 GHz sources is random.  Since there are 17 submm source (S/N $>4$), 
$\sim0.3$ and 0.7 random radio sources will be found within  $6\arcsec$ and 
$9\arcsec$ from the submm source positions, respectively.  In addition to the above
surface density analysis, we repeated a large number of simulations in which we
randomly shifted the radio sample and measured the submm fluxes at the shifted radio
positions.  On average, 0.9 radio sources per simulation have submm fluxes
greater than 3 $\sigma$, consistent with the surface density analysis.  Thus, there is 
a good chance that $\sim1$ of the 14 radio sources identified here is a 
chance projection.  However, most of the radio sources within the 
$\sim6\arcsec$ submm error circles are very likely connected to the submm 
sources in various ways.  Most are probably the real counterparts of the 
submm sources, but they could also be sources in the same groups as the 
submm sources, or sources lensed by the same foreground objects.

If we assume that the radio sources are the true counterparts of the submm 
sources, we can crudely estimate the redshifts of the submm sources using 
their submm to radio SEDs \citep{yun02} or simply their submm-to-radio flux 
ratios \citep[][Barger et al.\ 2000;]{carilli99}.  Using this millimetric redshift
technique, we estimated the redshifts of the 14 submm sources described above.  We
used the formula
\begin{equation}\label{z_mm}
z+1 = 0.98(S_{850}/S_{1.4})^{0.26}, 
\end{equation}
which was derived using the Arp 220 template (Barger et al.\ 2000).  The 850 
$\mu$m fluxes are the beam-optimized fluxes in Table~\ref{tab1}.  The results 
are listed in Table~\ref{tab3}.  For the two radio sources associated with submm
source GOODS 850-11, we assumed that only one radio source is responsible for the 
submm emission and calculated its submm-to-radio flux ratio.  If both radio
sources are responsible for the submm emission, then the redshifts in 
Table~\ref{tab3} are overestimated.  We also discuss GOODS 850-11 in
\S~\ref{individual}.  The errors in the redshifts 
in Table~\ref{tab3} only account for the errors in the 850 $\mu$m and 1.4 
GHz fluxes.  The actual uncertainty of this redshift estimate comes from 
the uncertainty in the submm-to-radio SED model and could be larger than 
0.5 \citep[see, e.g.,][]{ivison02}.

\subsection{X-Ray Counterpart Candidates}\label{x_id}
We searched for X-ray counterparts to the 850 $\mu$m sources using the 
CDF-N 2 Ms point-source catalog of 
\citet{alexander03}.  We used a searching method identical to that for the 
radio identifications, but we excluded X-ray sources that are $>10\arcsec$ 
away from the nearest submm source.  This is because the density of X-ray 
sources at the center of the ACS field is considerably higher than the
density of radio sources, and the probability of chance projections is 
higher.  Twenty X-ray sources were found in this way.  Fifteen X-ray sources 
are associated with 10 (out of 17) submm sources with S/N$>4$.  Three submm 
sources (GOODS 850-7, 11, and 13) have multiple X-ray sources associated with 
them.  Eight of the X-ray sources are associated with radio sources, all 
with positional offsets less than $0\farcs6$ between the radio and X-ray 
positions (see Figure~\ref{thumbnails}). The submm detected X-rays sources 
are summarized in Table~\ref{tab4} and Figure~\ref{thumbnails}.

The mean X-ray source surface density in the central $5\arcmin$ of the 
\emph{Chandra} field is $\sim8\times10^3$ deg$^{-2}$, corresponding 
to $\sim0.19$ random sources per $10\arcsec$ radius circle.  Because there are 
17 4 $\sigma$ submm sources searched, the above possibility suggests that
$\sim3$ of the 15 X-ray sources around the 4-$\sigma$ submm sources are
chance projections.  In addition, by randomly shifting the X-ray sample, we find that
the mean number of submm detected random X-ray sources is 2.7, consistent with the 
surface density analysis.  The fraction of bright submm sources (with $S_{850}$
$>6$ mJy and S/N $>4$) with X-ray counterparts is $\sim50\%$.  However, if the
number of random sources is taken into account, the fraction decreases to 
$\sim35\%$.  Thus radio sources are generally more likely to provide an astrometric
measurement of the submm source position than X-ray sources and much less likely to 
provide a spurious identification.

\subsection{MIR Counterpart Candidates}\label{mir_id}

Using the above methods, we searched for MIR counterparts to the 850 
$\mu$m sources using the main catalog in \citet{aussel99}, which made use 
of the Infrared Space Observatory 6.75 and 15 $\mu$m data.  Six MIR sources 
are found to have 850 $\mu$m fluxes greater than 3 $\sigma$ and are 
associated with five submm sources (Table~\ref{tab5}).  All of them are 15 
$\mu$m sources and only one has a detected 6.75 $\mu$m flux.  The offsets 
between the MIR and submm positions are between $3\arcsec$ and $10\arcsec$.  
We also estimated the number of random MIR sources based on the 15 $\mu$m 
source number counts in \citet{aussel99}.  The surface density of 15 $\mu$m 
sources brighter than 100 $\mu$Jy is $\sim7\times10^3$ deg$^{-2}$, 
corresponding to 0.17 sources per $10\arcsec$ radius circle.  There are $\sim10$ 
850 $\mu$m sources within the MIR field-of-view.  Thus we expect that one 
or two of the six MIR counterpart candidates are chance projections.  By 
randomly shifting the MIR sample, we find that the mean number of submm 
detected random MIR sources is 1.5, consistent with the surface density analysis.
Despite the comparable number counts, the fraction of chance 
projections for MIR sources is much higher than that for X-ray sources because 
the MIR observations are around the HDF-proper, which has the highest density 
of detected 850 $\mu$m sources.  We note that all of the MIR sources with redshifts
in Table~\ref{tab5} are at $z<1.0$, significantly lower than the millimetric or
photometric redshifts of the submm sources (see \S~\ref{individual} and
Table~\ref{tab7}).  We discuss these individual sources in \S~\ref{individual}.

\section{Optical Counterpart Candidates and Redshifts}

A ground-based, wide field, deep multi-color imaging survey centered on the HDF-N 
was conducted by \citet{capak04a}.  Accurate photometry and astrometry data 
in the $U$, $B$, $V$, $R$, $I$, and $z^{\prime}$ bands covering 0.2 deg$^2$, and 
additional $HK^{\prime}$ band data over a smaller region covering the CDF-N are
available.  Using these data, \citet{capak04b} derived photometric redshifts of the
galaxies in the HDF, based on the Bayesian technique in \citet{benitez00}.  
Here we use the photometry and photometric redshift data for
an overview of the 850 $\mu$m source counterpart candidates, as well as the radio,
X-ray, and MIR sources mentioned in the previous section.  Because many of the 
galaxies around the submm sources are optically faint, their photometric redshifts
have large uncertainties.  We limit ourselves to photometric redshifts that have  
$>80\%$ confidence, i.e., the probability that the $95\%$ errors are correct is greater 
than 80\% (see \citealp{benitez00} and \citealp{capak04b}).  Spectroscopic redshift 
data are also used, if available.    The photometric redshifts are especially useful 
when the galaxies are also radio sources.  If a photometric redshift coincides with 
the millimetric redshift, the identification may be more secure.  Table~\ref{tab6} 
lists the optical and NIR
magnitudes and redshifts of selected galaxies that are inside the positional error 
circles of the submm sources or that are associated with the radio,  X-ray, and MIR
sources.  Galaxies in Table~\ref{tab6} are labeled in Fig.~\ref{thumbnails}.  In
Fig.~\ref{thumbnails}, we use the HST/ACS images \citep{giavalisco04} for their
higher resolution.  All magnitudes in this paper are in the AB system, where 
$m_{\rm AB}=8.90-2.5\mathrm{log}_{10}(S/\rm Jy)$.  

\subsection{Individual Sources}\label{individual}

Now we briefly comment on each of the 4 $\sigma$ sources.  The fluxes quoted after
the source names are their 850 $\mu$m fluxes.

\begin{figure}[!h]
\epsscale{1.05}
\plottwo{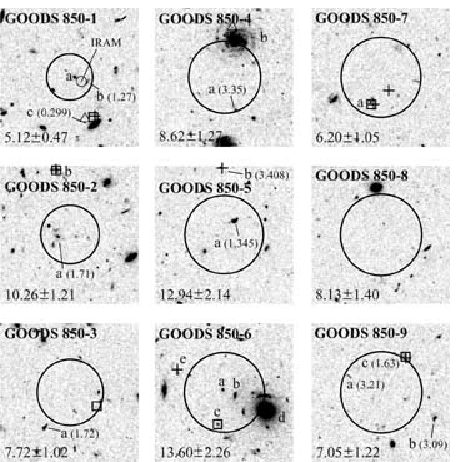}{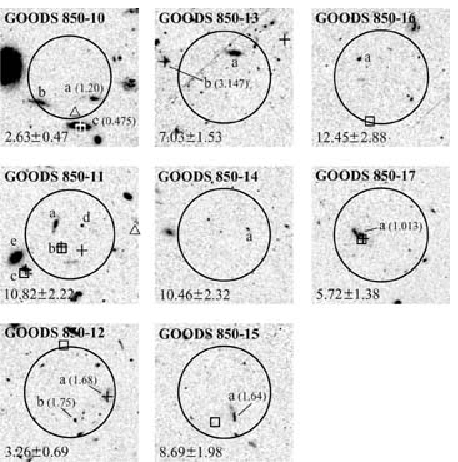}
\caption{HST ACS thumbnail images for the $4\sigma$ sources in Table~\protect\ref{tab1}.  
The grayscale images are summed F435, F606, F775, and F850 images, centered on 
the submm positions.  The panels are $18\arcsec$ on a side.  North is up and 
east is to the left.  The submm flux and uncertainty of each source are labeled 
at the bottom of each panel.  Large circles in the panels are the error circles 
for the submm positions.  Approximately 84\% of the real counterparts to the submm 
sources should be within the error circles.  Squares, crosses, and triangles are 
respectively 1.4 GHz sources, X-ray sources, and 15 $\mu$m sources with 3 $\sigma$ 
submm detections.  Optical spectroscopic redshifts (3 significant figures) or reliable 
($>80\%$ confidence) photometric redshifts (2 significant figures) are labeled.  To 
avoid confusing the plot, we only label photometric redshifts of radio and X-ray 
sources, and galaxies with $z>1.0$.  See Table~\protect\ref{tab6} for a full list 
of available redshifts. \label{thumbnails}}
\end{figure}


\bf GOODS 850-1~\rm ($5.12 \pm 0.47$ mJy)~
This is the brightest submm source in the HDF-proper region and is among the best
studied submm sources.  We refer to \citet{downes99} and \citet{serjeant03} for 
more detailed discussions of the identification of this source.  An IRAM  
interferometric observation at 1.3 mm had located the dust continuum between 
the optical galaxies 1a and 1b \citep{downes99}.  This rules out the radio and X-ray
source (and perhaps the MIR source, too) 1c at $z=0.300$ \citep{wirth04}
as the counterpart to the submm source.  The IRAM position coincides with a 4.5
$\sigma$ detection at 8.5 GHz \citep{richards98} and further strengthens the
identification.  The red galaxy 1a has a photometric redshift of 0.93.  The galaxy 1b
has a photometric redshift 1.27 and has an arc-like morphology, suggesting that it
may be lensed by 1a \citep{downes99,serjeant03}.  \citet{dunlop02} claimed that the
submm source is a faint object behind the galaxy 1a at $z\sim4$.  Using 
Eq.~\ref{z_mm} and the 40 $\mu$Jy detection limit at 1.4 GHz, the non-detection at
1.4 GHz implies a redshift lower limit of $\sim2.4$, consistent with a similar estimate
made by \citet{serjeant03} using the 8.5 GHz flux.  This lower limit does not 
completely rule out 1b as the counterpart to the submm emission because of the
uncertainties in the millimetric and photometric redshifts.  

\bf GOODS 850-2~\rm ($10.26 \pm 1.21$ mJy)~
A radio and X-ray source (2b) could be associated with the submm source.  
However, the offset between the radio and submm positions is $8\farcs6$, more than
2 times greater than the positional error corresponding to the submm S/N.   
Thus 2b is not likely to be the counterpart to the submm source.  Another galaxy (2a)
within the error circle has a photometric redshift of $1.71^{+0.5}_{-0.4}$ and a fairly
flat spectrum from $U$ to $z^{\prime}$, and perhaps $HK^{\prime}$, consistent
with an irregular or starburst galaxy at $z>1.3$.  Indeed, the HST/ACS image shows 
an irregular morphology and fuzzy light around the galaxy, which might be tidal features.  
Since 2a is closer to the submm center and has a starburst and interaction 
signature, it 
could be a better counterpart candidate than 2b.  However, if 2a were the real 
starbursting counterpart at $z=1.7$ and Arp 220 like, its 1.4 GHz flux would be
$\gtrsim200$ $\mu$Jy, which would have been detected easily.  If 2b is not the real
counterpart to GOODS 850-2, then its non-detection in the radio suggests that the
lower limit of its redshift is $\sim3.0$.

\bf GOODS 850-3~\rm ($7.72 \pm 1.02$ mJy)~
A radio source could be associated with GOODS 850-3 with a millimetric redshift of
1.73.  An optical galaxy (3a) with irregular morphology, a flat $U$ to $z^{\prime}$
spectrum, and a photometric redshift of 1.72 is another possibility.   Approximately
$1\farcs2$ to the northwest of the radio source, there are three galaxies (not labeled in
Figure~\ref{thumbnails}) in the ACS image that appear to be interacting with each other.  
These galaxies are not resolved in the ground-based images, and their combined optical SED 
has a photometric redshift of $1.25^{+0.45}_{-47}$ with confidence $80\%$.  
This photometric redshift is consistent with the millimetric redshift but is reliable only
if the three galaxies are at the same redshift.  If this is the case, then the interacting
group of galaxies and the radio source may be the source of the submm emission.

\bf GOODS 850-4~\rm ($8.62 \pm 1.27$ mJy)~
A MIR source is associated with a $V=22.83$ regular spiral galaxy at $z=0.848$
(4b, \citealp{wirth04}).  If the submm emission came from the star formation in this
spiral galaxy and the galaxy obeyed the FIR--radio correlation, the corresponding 1.4
GHz emission would have been detected.  In addition, the 8.62 mJy 850 $\mu$m flux
of  GOODS 850-4 corresponds to an ultraluminous $\lesssim10^{13}$ $L_{\sun}$
infrared luminosity.  It is highly unlikely that an ultraluminous infrared galaxy has an
undisturbed spiral morphology. There is a fainter galaxy (4a) inside the error circle
with a photometric redshift of 3.35.   This redshift is consistent with the non-detection
in the radio.

\bf GOODS 850-5~\rm ($12.94 \pm 2.14$ mJy)~
The brightest optical galaxy inside the error circle (5a) has a spectroscopic redshift 
of 1.345 \citep{wirth04}.  If this galaxy were the optical counterpart to the submm
source, its 1.4 GHz flux would be $>400$ $\mu$Jy.  The non-detection in the radio
for GOODS 850-5 sets the lower limit of the redshift to be $\sim3.4$.  An X-ray
source (5b) $8\farcs7$ to the north of the submm position has a spectroscopic redshift
of 3.408 \citep{cohen00}, consistent with the above lower limit.  The galaxy 5b and
the X-ray source could be the source of the submm emission.

\bf GOODS 850-6~\rm ($13.60 \pm 2.26$ mJy)~
This submm source is the brightest one in our sample.  The $V=22.17$ galaxy 6d 
has a spectroscopic redshift of 0.512 \citep{cowie04}.   
Two galaxies (not labeled) around 6d have morphologies similar to lensed arcs.  
The galaxy 6d could be a gravitational lens.  This lens candidate is also
noted in \citet{moustakas04}.   The galaxy 6a has a red color and a
photometric redshift 0.74.  A radio source is associated with 6c, which does not 
have a reliable photometric redshift but has a millimetric redshift of 2.45.  
Galaxy 6b has a photometric redshift of 0.14 and thus is less likely the counterpart 
to the submm source.

\bf GOODS 850-7~\rm ($6.20 \pm 1.05$ mJy)~
Two X-ray sources and one radio source are associated with this submm source.  The 
source 7a was detected in the ACS and ground-based image, but is too faint for a 
reliable photometric redshift.  If the radio source is the counterpart to the submm
source, the millimetric redshift is 2.37.

\bf GOODS 850-8~\rm ($8.13 \pm 1.40$ mJy)~
No optical counterpart candidates could be identified within the error circle.  The only 
bright galaxy slightly to the north of the error circle (not labeled) has a photometric 
redshift of 0.10 and is not likely the counterpart to the submm source.  The 
non-detection in the radio sets a redshift lower limit of 2.9.

\bf GOODS 850-9~\rm ($7.05 \pm 1.22$ mJy)~
Two galaxies near the error circle (9a and 9b) have photometric redshifts $>3.0$.  
If either of these two galaxies is the submm source, the $>3$ redshift will be consistent 
with the non-detection in the radio.  The galaxy 9c is a radio and X-ray
source and has a photometric redshift of 1.63.  The millimetric redshift of 9c is 2.64.  

\bf GOODS 850-10~\rm ($2.63 \pm 0.47$ mJy)~
The radio, X-ray, and MIR sources appear to be associated with the galaxy 10c,
which has a spectroscopic redshift of 0.475 \citep{wirth04}.   Because of its low
redshift, the galaxy 10c may not be the counterpart to the submm source.   In addition,
the millimetric redshift derived from the submm flux and the radio flux is 1.76.   
Thus, it is very likely that the radio source is not associated with the submm source
and the millimetric redshift is wrong.  There are two other bright galaxies, 10a and
10b, inside the error circle with redshifts of 1.20 and 0.851 \citep{cohen00},
respectively.  If any of these $z\sim1.0$ galaxies were the submm source, its 1.4 GHz
flux would be $>100$ $\mu$Jy and would have been detected.  The lack of 1.4 GHz
detection sets a lower limit of 1.9 to the millimetric redshift.  A similar problem was
noted by \citet{serjeant03}.  The lack of a radio detection for 10a and 10b, and the
inconsistency between the spectroscopic and millimetric redshifts of 10c leave 
no obvious identification for GOODS 850-10.

\bf GOODS 850-11~\rm  ($10.82 \pm 2.22$ mJy)~
Its submm flux is among the higher ones in the HDF, and there is a high density of
sources in this region.  Two radio sources, three X-ray sources, and two MIR sources
are associated with GOODS 850-11.   Two optically bright galaxies (11a and 11e) 
are at redshifts of 0.680 \citep{wirth04} and 0.556 \citep{cowie04}.   Galaxy 11e 
appears to be the closest counterpart candidate to one MIR source whose MIR position is 
slightly outside the figure.  All other radio, X-ray, and MIR sources are
either not detected in the optical or are too faint to obtain
reliable photometric redshifts.  These sources, as well as GOODS 850-11, might be in
the same group as 11a or 11e, but they are more likely at higher redshifts.  
Millimetric redshifts based on either of the two radio sources (11b and 11c) are 2.13
and 2.18 (\S~\ref{radio_id}).  If both radio sources are in the same group and are Arp
220-like, their redshifts will be 1.64, based on the total 1.4 GHz and submm
fluxes, and their physical separation will be $\sim50$ kpc ($H_0=71$ km s$^{-1}$
Mpc$^{-1}$, $\Omega_M=0.73$, $\Omega_{\Lambda}=0.27$).  Either redshift 
(1.6 or 2.1) is significantly higher than the redshifts of 11a and 11e, suggesting that
galaxies 11a and 11e could act as lenses on the background submm, radio, X-ray, and
MIR sources.  However, the radio sources 11b and 11c are not likely multiple images
of one object because of their different optical colors (Table~\ref{tab6}).
We note that 11c appears to be the brighter member of an interacting pair 
of galaxies.

\bf GOODS 850-12~\rm ($3.26 \pm 0.69$ mJy)~
Two galaxies inside the error circle appear to have similar photometric redshifts of
$\sim1.7$.  The galaxy 12a is resolved in the ACS image and has an irregular
morphology.  An optically faint radio source is $6\farcs2$ away from the submm
position.  If this radio source is the counterpart to the submm source, its millimetric
redshift is 1.96.

\bf GOODS 850-13~\rm ($7.03 \pm1.53 $ mJy)~ 
Two X-ray sources are associated with GOODS 850-13.  The source 13b has a 
spectroscopic redshift of 3.147 \citep{barger03} and has a red NIR to optical color.  
The other X-ray source is optically faint.  The brighter galaxy (13a) has a redshift of
0.556 \citep{wirth04} and is less likely to be the counterpart to the submm source
because of its low redshift.

\bf GOODS 850-14~\rm ($10.46 \pm 2.32$ mJy)~
This is an optically faint submm source.  The brightest galaxy inside the error 
circle (14a) has $V=26.18$ and has no reliable photometric redshift.  The bright galaxy 
outside the error circle (not labeled) has a photometric redshift of 0.12, and
therefore is less likely to be the source of the submm emission.  The non-detection in
the radio sets a lower limit of 3.2 to its millimetric redshift.

\bf GOODS 850-15~\rm ($8.69 \pm 1.98$ mJy)~
The radio source associated with GOODS 850-15 is optically faint and has a millimetric 
redshift of 1.83.  The radio source is marginally resolved by the
interferometer and has a $2\farcs5$ FWHM \citep{richards00}.  A galaxy (15a) with
a photometric redshift of 1.64 is $2\farcs6$ to the west of the radio source and may be
physically connected to the radio source.  This galaxy also shows signatures of
interaction with a faint galaxy at $\sim0\farcs8$ to its north.  In the radio map there is
a $\sim32$ mJy peak associated with this galaxy, which is slightly below the 40
$\mu$Jy detection limit and therefore not cataloged in \citet{richards00}.  This hint 
of radio emission suggests that the galaxy is a starbursting galaxy.  Given the
proximity of the radio source and galaxy 15a, and the coincident optical and
millimetric redshifts, galaxy 15a and the radio/submm source are likely to be in the
same group of galaxies at a redshift between 1.6 and 1.8.  

\bf GOODS 850-16~\rm ($12.45 \pm 2.88$ mJy)~
An optically faint radio source is associated with GOODS 850-16.  It has a millimetric 
redshift of 1.53.  Galaxy 16a is the brightest galaxy within the error circle
and has a photometric redshift of 0.70.  It is not likely to be the source of the submm 
emission.

\bf GOODS 850-17~\rm ($5.72 \pm 1.38$ mJy)~
The radio source and the X-ray source coincide with a pair of interacting galaxies 
at $z=1.013$ \citep{cohen00}.  This pair of galaxies is likely also the source of the
submm emission.  The millimetric redshift is 1.96.  The difference between the
spectroscopic redshift and the millimetric redshift is somewhat larger than we might
expect, indicating that this radio and submm source is not Arp 220-like.  

\subsection{Redshifts of the Sources}

In Table~\ref{tab7} we summarize the redshifts of the most plausible candidates 
or the redshift lower limits discussed above.  For sources with both millimetric and
photometric redshifts, we use the photometric ones given their relatively lower
uncertainties.  Excluding those with only lower limits, the median redshift of 11
possibly identified sources is 2.0.  Including the lower limits set by the radio
non-detections, the lower limit of the median redshift is 2.4.  This 2.4 lower limit is
consistent with \citet{ivison02} and \citet{chapman03b} but is considerably 
greater than the lower limit of 1.4 in \citet{webb03a,webb03b}.  The analyses of all
these groups, including ours,  intensively use radio images.  The 1.4 GHz radio
image used here and in Ivison et al.\ is significantly deeper than that in Webb et al., 
so Webb et al.\ can only detect radio emission from sources at lower redshifts 
(see e.g., their Table~6).  This may be the origin of the difference.  \citet{lilly99} 
presented eight submm sources and their spectroscopic and photometric redshifts.
Their sources are optically ($I$-band) identified based on positional coincidence.
Four of their sources have $z<1$ and the others have $1<z<3$.  Only two of their
sources have 5 GHz radio counterparts.  The significant fraction of their sources at
$z<1$ is very different from the redshifts measured by all the other groups mentioned
above.  This may be due to misidentifications caused by the higher surface density of
optically bright galaxies at lower redshifts. 

The above comparison clearly illustrates the limitation of our current observational 
techniques.  While pure optical identifications are biased toward low redshift
($z\lesssim 1$) and  optically bright galaxies, radio identifications can accurately
locate the submm emission and  are capable of finding optically faint submm sources
at much higher redshifts ($z\lesssim2.5$).  On the other hand, the sensitivities of
current radio interferometers still limit the highest redshift that we can reach.  To
identify sources at even higher redshifts, we will need either more sensitive radio 
observations to detect the redshifted nonthermal emission, or submm interferometric
imaging to directly locate the submm emission.  The latter is now becoming possible
with the advent of the Submillimeter Array \citep{moran98}.

\section{Summary}
1. We have carried out an 850 $\mu$m survey with SCUBA on the JCMT of 
an $\sim110$ arcmin$^2$ area centered on the HDF-N, with $1\sigma$ sensitivities
of 0.4 to 4 mJy.  Our source catalog is fully consistent with previous jiggle-map 
studies.  However, there is a serious discrepancy between out jiggle map
and the scan map of Borys et al.\ (2002).  There is also marginal
inconsistency between our jiggle map and photometry studies of \citet{chapman01} and 
\citet{chapman03b}.

2. After taking into account the effects of noise, confusion, incompleteness, and the
Eddington bias using Monte Carlo simulations, we find that the observed cumulative
850 $\mu$m source number counts between 2 and 10 mJy are consistent with a 
single power 
law $N(>S) = 1.09  \times 10^4 (S/ \rm mJy)^{-1.76}$ deg$^{-2}$.  Our number counts
are also consistent with previous measurements in blank fields and lensing cluster
fields.  In the 2-10 mJy flux range, the integrated submm source surface brightness
accounts for $34^{+20}_{-11}\%$ or $24^{+14}_{-7}\%$ of the FIR EBL,
depending on which measurement of the FIR EBL is adopted.

3. Radio, X-ray, and MIR counterpart candidates are identified near the submm
source positions.  Using surface density analyses and Monte Carlo simulations, the 
numbers of random sources around the submm sources are estimated.  The results
suggest that most of these counterpart candidates are physically connected to the 
submm sources and are not chance projections along the line of sight.  

4. The radio counterparts are used to estimate the redshifts of the submm sources by
assuming the Arp 220 SED.  Optical counterpart candidates are also selected around 
the submm sources or at the radio and X-ray positions.  Spectroscopic redshifts 
and photometric redshifts are also used, if available.  For 11 possibly identified
sources, the median redshift is 2.0.  Using the lower limits provided by the
radio non-detections, we find a lower limit of 2.4 for the median redshift
of 17 sources.
Identifications of submm sources at higher redshifts have to await the advent of 
high resolution submm imaging or more sensitive radio imaging.

\acknowledgments
We thank Peter Capak for useful discussions about the optical and NIR photometry
and for kindly providing us with the photometric redshift data in advance of
publication.  We thank the referee, Steve Eales, for the comprehensive report
that improved the manuscript, and Colin Borys for communicating with us about 
the SCUBA noise artifact and the JCMT tracking error.  
W.-H. W. and L. L. C. gratefully acknowledge support from NASA
grants G02-3187B and HST-G0-09425.03-A and NSF grant AST-0084816.  
A. J. B. gratefully acknowledges support from NSF grants AST-0084847, 
AST-0239425, NASA grant HST-G0-09425.30-A,
the University of Wisconsin Research Committee with founds granted by the 
Wisconsin Alumni Research Foundation, the Alfred P. Sloan Foundation, and the
David and Lucile Packard Foundation.

\begin{deluxetable}{lrrrrrr}
\tablecaption{850 $\mu$m Sources in the HDF.  \label{tab1}}
\tablewidth{0pt}
\tabletypesize{\footnotesize}
\tablehead{\colhead{ID}  &
\colhead{R.A. (2000.0)\tablenotemark{1}}  &
\colhead{Dec. (2000.0)\tablenotemark{1}}  &
\colhead{$S_{850}$ (mJy)\tablenotemark{1}} &
\colhead{S/N}  &
\colhead{$S_{ap}$ (mJy)\tablenotemark{2}} &
\colhead{$r_e$\tablenotemark{3}}  }
\startdata
 GOODS 850-1 & 12 36 52.20 & 62 12 26.50 & $ 5.12 \pm 0.47$ & 10.93 &  4.51 &  $3\farcs0$\\
 GOODS 850-2 & 12 36 22.40 & 62 16 21.31 & $10.26 \pm 1.21$ &  8.50 & 10.91 &  $3\farcs8$\\
 GOODS 850-3 & 12 36 18.83 & 62 15 52.26 & $ 7.72 \pm 1.02$ &  7.59 &  8.37 &  $4\farcs3$\\
 GOODS 850-4 & 12 36 37.05 & 62 12 08.45 & $ 8.62 \pm 1.27$ &  6.76 &  5.54 &  $4\farcs7$\\
 GOODS 850-5 & 12 36 33.45 & 62 14 09.43 & $12.94 \pm 2.14$ &  6.04 & 18.52 &  $5\farcs1$\\
 GOODS 850-6 & 12 37 30.68 & 62 13 03.15 & $13.60 \pm 2.26$ &  6.01 & 19.41 &  $5\farcs2$\\
 GOODS 850-7 & 12 36 15.97 & 62 15 17.22 & $ 6.20 \pm 1.05$ &  5.90 &  6.78 & $5\farcs2$ \\
 GOODS 850-8 & 12 36 06.30 & 62 12 47.05 & $ 8.13 \pm 1.40$ &  5.80 &  9.15 &  $5\farcs2$\\
 GOODS 850-9 & 12 37 07.66 & 62 14 03.44 & $ 7.05 \pm 1.22$ &  5.77 &  8.72 &  $5\farcs3$\\
 GOODS 850-10 & 12 36 49.91 & 62 13 19.50 & $ 2.63 \pm 0.47$ &  5.61 &  3.14 &  $5\farcs4$\\
 GOODS 850-11 & 12 36 45.90 & 62 14 50.49 & $10.82 \pm 2.22$ &  4.86 & 12.71 &  $5\farcs8$\\
 GOODS 850-12 & 12 36 56.49 & 62 12 01.49 & $ 3.26 \pm 0.69$ &  4.75 &  3.11 &  $5\farcs9$\\
 GOODS 850-13 & 12 37 13.21 & 62 12 06.39 & $ 7.03 \pm 1.53$ &  4.61 & 11.65 &  $6\farcs0$\\
 GOODS 850-14 & 12 36 23.45 & 62 13 16.33 & $10.46 \pm 2.32$ &  4.52 &  7.66 &  $6\farcs0$\\
 GOODS 850-15 & 12 36 21.10 & 62 17 12.30 & $ 8.69 \pm 1.98$ &  4.38 &  8.55 &  $6\farcs0$\\
 GOODS 850-16 & 12 37 00.05 & 62 09 15.48 & $12.45 \pm 2.88$ &  4.32 & 21.57 &  $6\farcs1$\\
 GOODS 850-17 & 12 36 28.77 & 62 10 46.39 & $ 5.72 \pm 1.38$ &  4.15 &  5.75 &  $6\farcs1$\\
\hline
 GOODS 850-18 & 12 36 10.73 & 62 12 34.13 & $ 4.80 \pm 1.22$ &  3.95 &  6.34 &  $6\farcs2$\\
 GOODS 850-19 & 12 36 36.61 & 62 12 42.45 & $ 3.26 \pm 0.85$ &  3.82 &  4.80 &  $6\farcs3$\\
 GOODS 850-20 & 12 37 13.99 & 62 16 33.38 & $ 7.14 \pm 1.87$ &  3.81 &  7.99 &  $6\farcs3$\\
 GOODS 850-21 & 12 35 53.65 & 62 10 15.75 & $ 7.22 \pm 1.91$ &  3.78 &  5.07 &  $6\farcs3$\\
 GOODS 850-22 & 12 36 15.18 & 62 12 00.21 & $ 4.19 \pm 1.15$ &  3.65 &  2.56 &  $6\farcs3$\\
 GOODS 850-23 & 12 37 14.21 & 62 11 50.38 & $ 5.52 \pm 1.53$ &  3.60 &  5.47 & $6\farcs3$ \\
 GOODS 850-24 & 12 36 19.21 & 62 10 04.27 & $ 6.01 \pm 1.70$ &  3.54 &  5.80 &  $6\farcs4$\\
 GOODS 850-25 & 12 35 53.49 & 62 10 40.75 & $ 6.16 \pm 1.74$ &  3.54 &  3.54 & $6\farcs4$ \\
 GOODS 850-26 & 12 37 36.38 & 62 12 18.04 & $ 6.01 \pm 1.71$ &  3.51 &  4.36 & $6\farcs4$ \\
 GOODS 850-27 & 12 37 54.48 & 62 14 54.60 & $ 6.17 \pm 1.76$ &  3.50 &  4.72 &  $6\farcs4$\\
\hline
 GOODS 850-28 & 12 37 18.86 & 62 16 51.33 & $11.04 \pm 3.23$ &  3.42 &  6.98 &  $6\farcs4$\\
 GOODS 850-29 & 12 36 18.13 & 62 14 53.25 & $ 4.00 \pm 1.18$ &  3.40 &  4.09 & $6\farcs4$ \\
 GOODS 850-30 & 12 36 13.17 & 62 12 20.17 & $ 3.75 \pm 1.10$ &  3.39 &  2.55 &  $6\farcs4$\\
 GOODS 850-31 & 12 36  09.75 & 62 11 46.11 & $ 4.37 \pm 1.31$ &  3.35 &  4.62 &  $6\farcs4$\\
 GOODS 850-32 & 12 36 53.63 & 62 13 55.50 & $ 1.98 \pm 0.60$ &  3.31 &  2.06 &  $6\farcs4$\\
 GOODS 850-33 & 12 36  02.06 & 62 10 44.96 & $ 5.38 \pm 1.63$ &  3.30 &  2.99 &  $6\farcs4$\\
 GOODS 850-34 & 12 36  07.22 & 62 15 54.07 & $ 5.96 \pm 1.81$ &  3.30 &  1.96 &  $6\farcs4$\\
 GOODS 850-35 & 12 37 10.83 & 62 16  08.41 & $ 5.77 \pm 1.78$ &  3.23 &  8.84 &  $6\farcs5$\\
 GOODS 850-36 & 12 37 11.66 & 62 13 27.41 & $ 4.40 \pm 1.36$ &  3.23 &  6.26 &  $6\farcs5$\\
 GOODS 850-37 & 12 37  09.47 & 62  9 17.42 & $11.26 \pm 3.50$ &  3.22 & 12.69 &  $6\farcs5$\\
 GOODS 850-38 & 12 36 25.19 & 62 11 50.35 & $ 6.11 \pm 1.93$ &  3.17 & 11.20 & $6\farcs5$ \\
 GOODS 850-39 & 12 38  00.57 & 62 13  00.42 & $ 7.41 \pm 2.36$ &  3.14 &  7.28 & $6\farcs5$\\
 GOODS 850-40 & 12 36  09.73 & 62 12 55.11 & $ 3.90 \pm 1.26$ &  3.10 &  6.43 & $6\farcs5$\\
 GOODS 850-41 & 12 36 57.66 & 62 19 27.49 & $ 4.38 \pm 1.41$ &  3.10 &  2.10 & $6\farcs5$\\
 GOODS 850-42 & 12 36 34.35 & 62  9 41.44 & $ 4.23 \pm 1.38$ &  3.06 &  3.40 & $6\farcs5$\\
 GOODS 850-43 & 12 37 21.09 & 62 13  01.30 & $ 5.53 \pm 1.83$ &  3.02 &  6.41 & 6$\farcs5$\\
 GOODS 850-44 & 12 36  08.06 & 62 16 55.08 & $ 7.69 \pm 2.56$ &  3.00 &  4.02 &$6\farcs5$ \\
 GOODS 850-45 & 12 37  08.80 & 62 13 48.43 & $ 3.57 \pm 1.19$ &  3.00 &  6.41 & $6\farcs5$\\
\enddata
\tablenotetext{1}{Beam-optimized source coordinates, fluxes, and errors that satisfy minimum $\chi^2$ for individual sources. See \S~\ref{direct_extract}.}
\tablenotetext{2}{$30\arcsec$ diameter aperture fluxes measured at the beam-optimized positions.  Extended sources or multiple sources resolved by the telescope have aperture fluxes significantly greater than the beam-optimized fluxes.}
\tablenotetext{3}{Position error associated with each source, see \S~\ref{counterparts}.}
\end{deluxetable}

\begin{deluxetable}{crrlrr}
\tablecaption{Scan-Map Sources and Their Jiggle-Map Counterparts.  \label{tab2}}
\tablewidth{0pt}
\tabletypesize{\footnotesize}
\tablehead{\colhead{HDFSMM\tablenotemark{1}}  &
\colhead{$S_{\rm Scan}$ (mJy)}  &
\colhead{$S_{\rm Jiggle}$\tablenotemark{2} (mJy)}  &
\colhead{ID$_{\rm Jiggle}$\tablenotemark{3}} &
\colhead{$S_{\rm Jiggle}$\tablenotemark{4} (mJy)}  &
\colhead{$d$\tablenotemark{5}}  }
\startdata
3606$+$1138 & $15.4\pm3.4$ & $-1.41\pm1.36$ & GOODS 850-31 & $4.37\pm1.31$ & $24\farcs8$ \\
3608$+$1246 & $13.8\pm3.3$ & $5.83\pm1.47$ & GOODS 850-8& $8.13\pm 1.40$& $10\farcs5$ \\
&&& GOODS 850-40& $3.90\pm1.26$ & $16\farcs3$ \\
&&& GOODS 850-18 & $4.80\pm1.22$ & $23\farcs7$ \\
3611$+$1211 & $12.2\pm3.0$ & $-2.46\pm1.20$ & GOODS 850-18 & $4.80\pm1.22$ & $23\farcs1$ \\
&&& GOODS 850-30 & $3.75\pm1.10$ & $20\farcs2$ \\
&&& GOODS 850-31 & $4.37\pm1.31$ & $25\farcs6$ \\
3620$+$1701 & $13.2\pm2.9$ & $4.24\pm1.39$ & GOODS 850-15 & $8.69\pm1.98$ & $12\farcs6$ \\
3621$+$1250 & $11.4\pm2.8$ & $2.70\pm1.82$ & \nodata & \nodata & \nodata \\
3730$+$1051 & $14.3\pm3.2$ & \nodata & \nodata & \nodata & \nodata \\
3623$+$1016 & $10.3\pm2.9$ & $-0.22\pm1.40$ &  \nodata & \nodata & \nodata \\
3624$+$1746 & $12.6\pm3.4$ & $-0.64\pm3.04$ & \nodata & \nodata & \nodata \\
3644$+$1452 & $11.4\pm2.9$ & $8.12\pm2.39$ & GOODS 850-11 & $10.82\pm2.22$& $9\farcs9$ \\
3700$+$1438 & $10.1\pm2.9$ & $3.79\pm2.46$ & \nodata & \nodata & \nodata \\
3732$+$1606 & $12.1\pm3.3$ & \nodata               & \nodata & \nodata & \nodata \\
3735$+$1423 & $13.4\pm3.8$ & $0.79\pm1.96$ & \nodata & \nodata & \nodata \\
\enddata
\tablenotetext{1}{Scan-map source ID in Table~1 of \citet{borys02}.}
\tablenotetext{2}{Jiggle-map flux measured at the scan-map source position.}
\tablenotetext{3}{Source numbers (as in Table~\ref{tab1}) of jiggle-map sources within $30\arcsec$ of the scan-map position.}
\tablenotetext{4}{Jiggle-map flux for the nearby jiggle-map source.}
\tablenotetext{5}{Offset between the scan-map source and the nearby jiggle-map source.}
\end{deluxetable}

\begin{deluxetable}{rrrlrrc}
\tablewidth{0pt}
\tabletypesize{\footnotesize}
\tablecaption{1.4 GHz Counterpart Candidates \label{tab3}}
\tablehead{\colhead{R.A. (2000.0)\tablenotemark{1}} &
\colhead{Dec. (2000.0)\tablenotemark{1}} &
\colhead{$S_{1.4}$ ($\mu$Jy)\tablenotemark{1}} &
\colhead{ID$_{\rm Submm}$\tablenotemark{2}} &
\colhead{$z$\tablenotemark{3}} &
\colhead{d\tablenotemark{4}}&
\colhead{Note}}
\startdata
12 36 51.76 & 62 12 21.30 & $ 49.3\pm 7.9$ &  GOODS 850-1 & $2.28\pm0.16$ & 6\farcs0 & \tablenotemark{*}\\
12 36 22.65 & 62 16 29.74 & $ 70.9\pm 8.7$ &  GOODS 850-2 & $2.57\pm0.16$ & 8\farcs6 & \tablenotemark{*}\\
12 36 18.33 & 62 15 50.48 & $151.0\pm11.0$ &  GOODS 850-3 & $1.73\pm0.11$ & 3\farcs9 \\
12 37 30.80 & 62 12 58.98 & $107.0\pm 9.6$ &  GOODS 850-6 & $2.45\pm0.17$ & 4\farcs3 \\
12 36 16.15 & 62 15 13.67 & $ 53.9\pm 8.4$ &  GOODS 850-7 & $2.37\pm0.20$ & 3\farcs8 \\
12 37  07.21 & 62 14  08.08 & $ 45.3\pm 7.9$ &  GOODS 850-9 & $2.64\pm0.23$ & 5\farcs6 \\
12 36 49.71 & 62 13 12.78 & $ 49.2\pm 7.9$ & GOODS 850-10 & $1.76\pm0.17$ & 6\farcs9 & \tablenotemark{*}\\
12 36 46.05 & 62 14 48.69 & $124.0\pm 9.8$ & GOODS 850-11 & $2.13\pm0.18$ & 2\farcs1 & \tablenotemark{*}\\
12 36 46.76 & 62 14 45.45 & $117.0\pm 9.6$ & GOODS 850-11 & $2.18\pm0.18$ & 7\farcs8 & \tablenotemark{*}\\
12 36 56.60 & 62 12  07.62 & $ 46.2\pm 7.9$ & GOODS 850-12 & $1.96\pm0.21$ & 6\farcs2 \\
12 36 21.27 & 62 17  08.40 & $148.0\pm11.0$ & GOODS 850-15 & $1.83\pm0.18$ & 4\farcs1 \\
12 37  00.26 & 62  09  09.75 & $324.0\pm18.0$ & GOODS 850-16 & $1.53\pm0.16$ & 5\farcs9 \\
12 36 29.13 & 62 10 45.79 & $ 81.4\pm 8.7$ & GOODS 850-17 & $1.96\pm0.20$ & 2\farcs6 \\
12 37 11.34 & 62 13 31.02 & $132.0\pm10.1$ & GOODS 850-36 & $1.44\pm0.20$ & 4\farcs2 \\
\enddata
\tablenotetext{1}{Source coordinates and 1.4 GHz fluxes from \citet{richards00}.}
\tablenotetext{2}{Associating 850 $\mu$m source.}
\tablenotetext{3}{Radio--submm redshifts.}
\tablenotetext{4}{Angular separation between the radio positions and the submm positions.}
\tablenotetext{*}{These radio sources may not be the real counterparts to the submm emission.  See discussion in \S~\ref{individual}.}
\end{deluxetable}

\begin{deluxetable}{rrrrlr}
\tablewidth{0pt}
\tabletypesize{\footnotesize}
\tablecaption{X-ray Counterpart Candidates \label{tab4}}
\tablehead{\colhead{R.A. (2000.0)\tablenotemark{1}} &
\colhead{Dec. (2000.0)\tablenotemark{1}} &
\colhead{FB\tablenotemark{2}} &
\colhead{BR\tablenotemark{3}} &
\colhead{ID$_{\rm Submm}$\tablenotemark{4}} &
\colhead{d\tablenotemark{5}}}
\startdata
12 36 51.73 & 62 12 21.40 & $ 2.94$ & $ 1.34$ & GOODS 850-1 &  $6\farcs1$ \\
12 36 22.66 & 62 16 29.80 & $ 1.02$ & $>4.39$ & GOODS 850-2 &  $8\farcs7$ \\
12 36 33.49 & 62 14 18.10 & $ 1.16$ & $ 0.41$ & GOODS 850-5 &  $8\farcs7$ \\
12 37 31.55 & 62 13 06.10 & $ 0.40$ & $ 0.71$ & GOODS 850-6 &  $6\farcs8$ \\
12 36 15.83 & 62 15 15.50 & $ 1.77$ & $ 5.22$ & GOODS 850-7 &  $2\farcs0$ \\
12 36 16.11 & 62 15 13.70 & $ 1.02$ & $ 0.85$ & GOODS 850-7 &  $3\farcs7$ \\
12 37  07.20 & 62 14  07.90 & $ 0.98$ & $ 1.62$ & GOODS 850-9 &  $5\farcs5$ \\
12 36 49.71 & 62 13 13.20 & $ 0.16$ & $<0.44$ & GOODS 850-10 &  $6\farcs4$ \\
12 36 45.68 & 62 14 48.40 & $ 0.08$ & $<1.00$ & GOODS 850-11 &  $2\farcs6$ \\
12 36 46.06 & 62 14 48.90 & $ 0.08$ & \nodata & GOODS 850-11 &  $1\farcs9$ \\
12 36 46.72 & 62 14 45.90 & $<0.09$ & $<1.06$ & GOODS 850-11 &  $7\farcs3$ \\
12 36 55.79 & 62 12  00.90 & $ 0.40$ & $ 0.48$ & GOODS 850-12 &  $4\farcs9$ \\
12 37 12.09 & 62 12 11.30 & $ 0.38$ & $ 2.00$ & GOODS 850-13 &  $9\farcs3$ \\
12 37 14.32 & 62 12  08.40 & $ 0.79$ & $ 0.52$ & GOODS 850-13 &  $8\farcs0$ \\
12 36 29.11 & 62 10 45.90 & $ 2.31$ & $ 2.05$ & GOODS 850-17 &  $2\farcs4$ \\
12 36 13.02 & 62 12 24.10 & $ 2.04$ & $ 2.24$ & GOODS 850-30 &  $4\farcs1$ \\
12 36  09.75 & 62 11 45.90 & $ 0.30$ & $<0.46$ & GOODS 850-31 &  $0\farcs2$ \\
12 36  06.70 & 62 15 50.70 & $ 1.69$ & $ 0.25$ & GOODS 850-34 &  $4\farcs9$ \\
12 37 11.38 & 62 13 30.80 & $ 0.56$ & $ 1.25$ & GOODS 850-36 &  $3\farcs9$ \\
12 36 34.48 & 62  09 41.80 & $ 0.57$ & $ 0.83$ & GOODS 850-42 &  $1\farcs0$ 
\enddata
\tablenotetext{1}{Source coordinates from \citet{alexander03}.}
\tablenotetext{2}{0.5-8.0 keV full band flux ($10^{-15}$ ergs cm$^{-2}$ s$^{-1}$).}
\tablenotetext{3}{(2.0-8.0 keV)/(0.5-2.0 keV) count ratio.}
\tablenotetext{4}{Associating 850 $\mu$m source.}
\tablenotetext{5}{Angular separation between the X-ray positions and the submm positions.}
\end{deluxetable}

\begin{deluxetable}{rrrrlrr}
\tablewidth{0pt}
\tabletypesize{\footnotesize}
\tablecaption{Mid-Infrared Counterpart Candidates \label{tab5}}
\tablehead{\colhead{R.A. (2000.0)\tablenotemark{1}} &
\colhead{Dec. (2000.0)\tablenotemark{1}} &
\colhead{LW2 ($\mu$Jy)\tablenotemark{2}} &
\colhead{LW3 ($\mu$Jy)\tablenotemark{3}} &
\colhead{ID$_{\rm Submm}$\tablenotemark{4}} &
\colhead{d\tablenotemark{5}} &
\colhead{z\tablenotemark{6}}}
\startdata
12 36 51.90 & 62 12 21.00 &  $<36$ &  $48^{+32}_{-09}$ &  GOODS 850-1 &$5\farcs9$ & 0.300$^{(a)}$ \\ 
12 36 36.90 & 62 12 15.00 & $<113$ & $202^{+58}_{-50}$ &  GOODS 850-4 &$6\farcs6$ & 0.848$^{(a)}$ \\ 
12 36 49.80 & 62 13 15.00 &  $136^{+68}_{-57}$ & $320^{+39}_{-62}$ & GOODS 850-10 &$4\farcs6$ & 0.475$^{(a)}$\\ 
12 36 44.70 & 62 14 51.00 & $<329$ & $105^{+94}_{-21}$ & GOODS 850-11 &$8\farcs4$ & \nodata \\ 
12 36 47.20 & 62 14 48.00 & $<179$ & $144^{+72}_{-47}$ & GOODS 850-11 &$9\farcs4$ & 0.556$^{(b)}$ \\ 
12 36 54.10 & 62 13 57.00 &  $<42$ &  $47^{+31}_{-09}$ & GOODS 850-32 &$3\farcs6$ & 0.850$^{(a)}$ \\ 
\enddata
\tablenotetext{1}{Source coordinates from Aussel et al.\ (1999).}
\tablenotetext{2}{6.75 $\mu$m flux.}
\tablenotetext{3}{15 $\mu$m flux.}
\tablenotetext{4}{Associating 850 $\mu$m source.}
\tablenotetext{5}{Angular separation between the MIR positions and the submm positions.}
\tablenotetext{6}{Redshifts of nearest galaxies within $3\arcsec$ from the MIR
positions.  Data are from \citet[][a]{wirth04}, and \citet[][b]{cowie04}}
\end{deluxetable}

\begin{deluxetable}{lrrcccccccrl}
\tablewidth{0pt}
\tablecaption{Optical Properties of the Counterpart Candidates  \label{tab6}}
\tabletypesize{\footnotesize}
\tablehead{\colhead{ID}  &
\colhead{R.A. (2000.0)}  &
\colhead{Dec. (2000.0)}  &
\colhead{$U$\tablenotemark{1}} &
\colhead{$B$\tablenotemark{1}}  &
\colhead{$V$\tablenotemark{1}} &
\colhead{$R$\tablenotemark{1}}  &
\colhead{$I$\tablenotemark{1}} &
\colhead{$z^{\prime}$\tablenotemark{1}} &
\colhead{$HK^{\prime}$\tablenotemark{1}} &
\colhead{$z$\tablenotemark{2}} &
\colhead{Note\tablenotemark{3}}
}
\startdata
GOODS 850-1a & 12 36 52.11 & 62 12 26.46 & \nodata & 26.95 & 26.33 & 25.18 & 24.15 & 23.41 & 21.80 &  $0.93^{+0.25}_{-0.25}$ \\
GOODS 850-1b & 12 36 51.90 & 62 12 25.82 & 27.64 & 26.26 & 26.14 & 25.94 & 25.05 & 24.82 & 23.21 &  $1.27^{+0.39}_{-0.47}$ \\
GOODS 850-1c & 12 36 51.73 & 62 12 20.42 & 24.16 & 23.51 & 22.59 & 22.03 & 21.60 & 21.41 & 20.61 &  0.300$^{(a)}$ & R? X? I?\\
GOODS 850-2a & 12 36 22.62 & 62 16 21.28 & 24.86 & 24.57 & 24.60 & 24.44 & 24.30 & 24.48 & 24.45 &  $1.71^{+0.54}_{-0.41}$ \\
GOODS 850-2b & 12 36 22.61 & 62 16 29.58 & 25.62 & 25.17 & 25.09 & 24.85 & 24.45 & 24.21 & 23.06 & \nodata & R, X\\
GOODS 850-3a & 12 36 19.30 & 62 15 47.78 & 24.98 & 24.97 & 25.17 & 25.08 & 24.93 & 25.02 & 25.32 & $1.72^{+0.36}_{-0.67}$  \\
GOODS 850-4a & 12 36 36.84 & 62 12  4.25 & \nodata & 26.27 & 25.81 & 25.23 & 24.92 & 24.45 & \nodata &  $3.35^{+0.57}_{-3.16}$ \\
GOODS 850-4b & 12 36 36.82 & 62 12 13.37 & 23.82 & 23.45 & 22.83 & 22.09 & 21.09 & 20.79 & 19.52 &  0.848$^{(a)}$ & I\\
GOODS 850-5a & 12 36 33.25 & 62 14 11.22 & 23.99 & 23.99 & 23.92 & 23.79 & 23.71 & 23.27 & 24.59 &  1.345$^{(a)}$ \\
GOODS 850-5b & 12 36 33.52 & 62 14 18.14 & 27.79 & 25.85 & 25.38 & 25.20 & 25.30 & 25.72 & 23.76 & 3.408$^{(d)}$ & X\\
GOODS 850-6a & 12 37 30.69 & 62 13  3.50 & \nodata & 26.69 & 26.11 & 25.51 & 24.29 & 23.94 & 23.50 &  $0.74^{+0.23}_{-0.23}$ \\
GOODS 850-6b & 12 37 30.47 & 62 13  2.84 & 25.76 & 24.40 & 23.93 & 23.54 & 23.25 & 23.15 & 22.52 &  $0.14^{+0.15}_{-0.14}$ \\
GOODS 850-6c & 12 37 30.77 & 62 12 58.90 & 26.12 & 26.10 & 25.65 & 25.51 & 25.58 & 24.77 & 24.83 & \nodata & R\\
GOODS 850-6d & 12 37 29.92 & 62 13  0.94 & 24.56 & 23.19 & 22.17 & 21.00 & 20.38 & 20.10 & 19.08 &  0.512$^{(b)}$ \\
GOODS 850-6e & 12 37 31.56 & 62 13  5.76 & \nodata & 27.20 & 26.39 & 25.75 & 25.80 & 25.31 & 24.89 & \nodata & X\\
GOODS 850-7a & 12 36 16.14 & 62 15 14.00 & 27.76 & 26.59 & 26.13 & 25.36 & 25.03 & 25.00 & 23.30 & \nodata & R, X\\
GOODS 850-9a & 12 37  8.26 & 62 14  5.49 & \nodata & 26.73 & 26.11 & 26.05 & 25.66 & 25.54 & \nodata &  $3.21^{+0.55}_{-3.08}$ \\
GOODS 850-9b & 12 37  6.66 & 62 14  0.16 & 27.34 & 25.63 & 24.92 & 24.56 & 24.42 & 24.56 & 23.69 &  $3.09^{+0.54}_{-2.93}$ \\
GOODS 850-9c & 12 37  7.21 & 62 14  8.15 & 27.53 & 26.77 & 26.36 & 25.63 & 25.51 & 24.69 & 22.36 &  $1.63^{+0.54}_{-0.86}$ & R, X\\
GOODS 850-10a & 12 36 49.45 & 62 13 16.70 & 24.64 & 24.44 & 24.31 & 23.89 & 23.43 & 22.84 & 21.98 & $ 1.20^{+0.29}_{-0.29}$ \\
GOODS 850-10b & 12 36 50.50 & 62 13 16.28 & 25.14 & 24.94 & 24.49 & 23.76 & 22.94 & 22.66 & 21.37 &  0.851$^{(d)}$ \\
GOODS 850-10c & 12 36 49.73 & 62 13 13.15 & 24.36 & 23.76 & 23.00 & 22.14 & 21.69 & 21.48 & 20.37 &  0.475$^{(a)}$ & R, X, I?\\
GOODS 850-11a & 12 36 46.16 & 62 14 51.77 & 24.97 & 24.72 & 24.47 & 23.88 & 23.47 & 23.46 & 22.80 &  0.680$^{(a)}$ \\
GOODS 850-11b & 12 36 46.07 & 62 14 48.89 & \nodata & \nodata & 26.58 & 26.19 & 26.78 & 25.29 & \nodata & \nodata & R, X\\
GOODS 850-11c & 12 36 46.75 & 62 14 46.15 & 25.40 & 25.50 & 25.43 & 25.13 & 24.94 & 24.79 & \nodata & \nodata & R, X\\
GOODS 850-11d & 12 36 45.65 & 62 14 51.57 & 26.24 & 25.70 & 25.59 & 25.60 & 25.64 & 25.57 & \nodata & \nodata \\
GOODS 850-11e & 12 36 46.90 & 62 14 47.47 & 24.29 & 23.92 & 22.97 & 21.87 & 21.13 & 20.90 & 20.07 &  0.556$^{(b)}$ \\
GOODS 850-12a & 12 36 55.81 & 62 12  0.94 & 25.61 & 24.98 & 24.82 & 24.67 & 24.47 & 24.02 & 22.32 &  $1.68^{+0.35}_{-0.35}$ & X\\
GOODS 850-12b & 12 36 56.34 & 62 11 57.94 & 25.47 & 25.09 & 24.96 & 24.83 & 24.84 & 24.81 & 23.35 &  $1.75^{+0.63}_{-0.38}$ \\
GOODS 850-13a & 12 37 13.02 & 62 12  9.54 & 24.44 & 24.36 & 24.07 & 23.35 & 23.09 & 23.05 & 23.06 &  0.556$^{(a)}$ \\
GOODS 850-13b & 12 37 14.31 & 62 12  8.60 & 25.20 & 24.38 & 24.17 & 23.97 & 23.75 & 23.41 & 21.96 &  3.147$^{(c)}$ & X\\
GOODS 850-14a & 12 36 22.98 & 62 13 17.01 & 26.69 & 26.44 & 26.18 & 25.91 & 26.31 & 25.65 & \nodata & \nodata \\
GOODS 850-15a & 12 36 20.92 & 62 17  9.45 & 25.10 & 24.71 & 24.65 & 24.42 & 24.05 & 23.96 & 21.95 &  $1.64^{+0.35}_{-0.35}$ \\
GOODS 850-16a & 12 37  0.45 & 62  9 17.74 & 25.90 & 25.81 & 25.59 & 24.88 & 24.55 & 24.25 & \nodata &  $0.70^{+0.22}_{-0.35}$ \\
GOODS 850-17a & 12 36 29.16 & 62 10 46.31 & 26.11 & 25.70 & 24.98 & 24.05 & 22.89 & 22.36 & 20.46 &  1.013$^{(d)}$ & R, X\\
\enddata
\tablenotetext{1}{All magnitudes are in the AB system.  The 5 $\sigma$ limiting
magnitudes for $U$, $B$, $V$, $R$, $I$, $z^{\prime}$ and $HK^{\prime}$ are
respectively 27.1, 26.9, 26.8, 26.6, 25.6, 25.4, and 22.1.}
\tablenotetext{2}{Spectroscopic redsifts (3 significant figures) or photometric redshifts (2 significant figures) with $>80\%$ confidence.  The errors of photometric
redshifts are 95\% errors (see \citealp{benitez00} and \citealp{capak04b}).  
(a): data from \citet{wirth04}, (b): data from \citet{cowie04},
(c): data from \citet{barger03}, (d): data from \citet{cohen00}.}
\tablenotetext{3}{R: 1.4 GHz sources; X: X-ray sources; I: 15 $\mu$m sources}
\end{deluxetable}

\begin{deluxetable}{lrr}
\tablewidth{300pt}
\tablecaption{Summary of Redshifts for Submm Sources.\label{tab7}}
\tabletypesize{\footnotesize}
\tablehead{\colhead{ID} &
\colhead{$z_{\rm mm}$\tablenotemark{1}} &
\colhead{$z_{\rm opt}$\tablenotemark{2}}
}
\startdata
GOODS 850-1 &   $>2.4$ &  \\
GOODS 850-2 &   $>3.0$ &  \\
GOODS 850-3 &   $1.73$ & $1.25$ \\
GOODS 850-4 &   $>3.0$ & $3.35$ \\
GOODS 850-5 &   $>3.4$ & $3.408$ \\
GOODS 850-6 &   $2.45$ &  \\
GOODS 850-7 &   $2.37$ &  \\
GOODS 850-8 &   $>2.9$ &  \\
GOODS 850-9 &   $2.64$ & $1.63$ \\
GOODS 850-10 & $>1.9$? & \\
GOODS 850-11 & $>1.6$ & $$ \\
GOODS 850-12 & $1.96$ & \\
GOODS 850-13 & $>2.8$ & 3.147 \\
GOODS 850-14 & $>3.2$ & \\
GOODS 850-15 & $1.83$ & $1.64$ \\
GOODS 850-16 & $1.53$ &  \\
GOODS 850-17 & $1.96$ & $1.013$
\enddata
\tablenotetext{1}{Millimetric redshifts derived from Eq.~\ref{z_mm}.  Lower limits
are estimated for submm sources that are not detected in radio.  Also see text.}
\tablenotetext{2}{Optical redshifts.  Redshifts with four significant figures are 
spectroscopic redshifts.  Redshifts with three significant figures are photometric 
redshifts with $>80\%$ confidence.}
\end{deluxetable}

\end{document}